\documentclass{JHEP3}
\usepackage[dvips]{epsfig}
\usepackage{graphicx,amssymb,amsmath,amsfonts,cite}
\usepackage{color}
\usepackage{bm}

\usepackage{xcolor}

\usepackage[english]{babel}
\newcommand{\cA}{{\cal A}}  
  \newcommand{\cD}{{\cal D}}
\newcommand{\cE}{{\cal E}}  \newcommand{\cF}{{\cal F}}

  \newcommand{\cL}{{\cal L}}
  \newcommand{\cN}{{\cal N}}
  \newcommand{\cP}{{\cal P}}
  
  \newcommand{\cT}{{\cal T}}

\newcommand{\be}{\begin{equation}} \newcommand{\ee}{\end{equation}}
\newcommand{\bea}{\begin{eqnarray}} \newcommand{\eea}{\end{eqnarray}}
\newcommand{\beann}{\begin{eqnarray*}}  \newcommand{\eeann}{\end{eqnarray*}}
\newcommand{\bfig}{\begin{figure}} \newcommand{\efig}{\end{figure}}
\newcommand{\ba}{\begin{array}} \newcommand{\ea}{\end{array}}
\newcommand{\bcen}{\begin{center}} \newcommand{\ecen}{\end{center}}
\newcommand{\btab}{\begin{tabular}} \newcommand{\etab}{\end{tabular}}

\newcommand{\vev}[1]{\left\langle{#1}\right\rangle}

\newtheorem{Proposition}{Proposition}[section]

\newtheorem{Theorem}{Theorem}[section]
\newtheorem{Lemma}{Lemma}[section]

\newcommand{\bp}{\begin{Proposition}}   \newcommand{\ep}{\end{Proposition}}
\newcommand{\bt}{\begin{Theorem}}   \newcommand{\et}{\end{Theorem}}
\newcommand{\bl}{\begin{Lemma}}     \newcommand{\el}{\end{Lemma}}
\newcommand{\bc}{\begin{Corolary}} \newcommand{\ec}{\end{Corolary}}

\definecolor{darkgreen}{rgb}{0.0,0.7,0.}

\title{Lifshitz Field Theories at Non-Zero Temperature, Hydrodynamics and Gravity}

\author{Carlos Hoyos, Bom Soo Kim and Yaron Oz\\
Raymond and Beverly Sackler School of
Physics and Astronomy, Tel-Aviv University, Tel-Aviv 69978, Israel\\
E-mail: \email{choyos,bskim,yaronoz@post.tau.ac.il}
}

\abstract{We consider a covariant formulation of field theories with Lifshitz scaling, and analyze the energy-momentum tensor and the scale symmetry Ward identity.
We derive the equation of state and the ideal Lifshitz hydrodynamics in agreement with arXiv:1304.7481, where they were determined by using thermodynamics and symmetry properties. We construct the charged ideal Lifshitz hydrodynamics in the generating functional framework as well as in the gravitational holographic dual description. At the first viscous order, an analysis of the entropy current reveals two additional transport coefficients (one dissipative and one dissipationless) compared to the neutral case, contributing to the charge current
and to the asymmetric part of the energy-momentum tensor.
}

\preprint{TAUP-2976/13}

\begin{document}

%%%%%%%%%%%%%%%%%%%%%%%%%%%%%%%%%%%%%%%%%%%%%%%%%%%%%%%%%%%
\section{Introduction}
%%%%%%%%%%%%%%%%%%%%%%%%%%%%%%%%%%%%%%%%%%%%%%%%%%%%%%%%%%%

A phase transition at zero temperature may occur as the ground state of a many-body system is changed by tuning an external parameter. The boundary between the two phases is a quantum critical point \cite{Coleman:2005,Sachdev:2011,2008NatPh...4..186G}, characterized by a `Lifshitz' scaling symmetry
\begin{equation} \label{LifshitzScaling}
t\to \lambda^z t, \ \ x^i\to \lambda x^i,
\end{equation}
where $t$ is time and $x^i$, $i=1,\dots, d$ are space coordinates. The number $z$ is the dynamic critical exponent. For an ordinary relativistic conformal field theory $z=1$, but in general systems its value can be arbitrary. Quantum critical points are believed to
underlie  the exotic properties of heavy fermion compounds and other materials including high $T_c$ superconductors. These materials have a metallic phase (dubbed as `strange metal') whose properties cannot be explained within the ordinary Landau-Fermi liquid theory. 

Simple examples where $z\neq 1$ are multicritical Lifshitz points \cite{Hornreich:1975}. In its simplest incarnation, the Lifshitz point is characterized by a free energy density for a scalar order parameter $\phi$ that takes the form
\begin{equation}
F(\phi)=a_2 \phi^2+a_4\phi^4+\cdots+c_1(\partial_i\phi)^2+c_2(\partial_i^2\phi)^2+\cdots.
\end{equation}
This is of the same form as the Ginzburg-Landau theory. For the Lifshitz point the coefficient $c_1\to 0$, so it is necessary to consider higher derivative terms. 
A simple model exhibiting the scaling properties of a Lifshitz point is that of a free scalar with an action
\begin{equation}\label{lifshitzaction}
S=\int dt\,d^d x\,\left[\frac{1}{2}(\partial_t\phi)^2-\frac{\kappa}{2z}((\partial_i^2)^{z/2}\phi)^2 \right].
\end{equation}
This action is invariant under the scaling transformation 
\begin{equation}
\phi\to \lambda^{(z-d)/2}\phi \ ,
\end{equation}
along with (\ref{LifshitzScaling}). This model, first considered in \cite{2004AnPhy.310..493A} for $z=2$, can be extended to include interactions respecting the scale symmetry \cite{Anselmi:2008ry,Fitzpatrick:2012ww}. Strongly coupled theories with Lifshitz scaling can be described by gravity duals using a generalization of the gauge/gravity correspondence, also known as holographic duality \cite{Kachru:2008yh}. In the gravity dual the Lifshitz scaling is realized as an isometry of the metric, see also \cite{Taylor:2008tg,Koroteev:2007yp}. Of special interest is that Lifshitz scaling emerges at large distances in finite density systems with a holographic description that have been proposed as models of strange metals \cite{Hartnoll:2009ns,Hartnoll:2010gu}.

Quantum critical points, however, are not accessible directly by experiments. Instead, their existence can be inferred from the properties of the system at finite temperature. For instance, for strange metals some quantities exhibit universal power-like behaviour such as the resistivity, which is linear in the temperature $\bm{\rho} \!\sim\! T$ \cite{PhysRevLett.59.1337,PhysRevLett.85.626,2013Sci...339..804B}. 
It is well known that systems with ordinary critical points behave hydrodynamically with transport coefficients, whose temperature dependence is determined by the scaling at the critical point \cite{Hohenberg:1977}. Quantum critical systems also have hydrodynamic descriptions, as has been shown more recently for conformal field theories at finite temperature \cite{Kovtun:2004de}, fermions at unitarity \cite{Cao:2011} and graphene \cite{Mueller:2008,Fritz:2008,Mueller:2009}. A similar hydrodynamic description has been suggested for strange metals based on the large scattering rate measured in experiments \cite{PhysRevLett.69.2411,Zaanen,2013Sci...339..804B}, and we proposed recently a concrete realization \cite{Hoyos:2013eza}.

We derived most properties of Lifshitz hydrodynamics using symmetry considerations, in particular we derived the equation of state from the `trace' Ward identity of the energy-momentum tensor associated to the scaling symmetry \cite{Hoyos:2013eza}
\begin{equation}
z T^\mu_{\ \nu} u_\mu u^\nu- T^\mu_{\ \nu}P_\mu^{\ \nu}=0,
\end{equation}
where $u^\mu$ is the velocity of the fluid $u_\mu u^\mu=\eta_{\mu\nu} u^\mu u^\nu \!=\!-1$ and $P^{\mu\nu}\!=\! \eta^{\mu\nu}+u^\mu u^\nu$, $ \mu, \nu = (t, i)$. In this paper we re-derive and extend some of our previous results from various perspectives. In addition to the dissipative transport coefficient $\alpha $ that we found in the neutral fluid \cite{Hoyos:2013eza}, we find two new transport coefficients $\alpha'$ and $C$ for charged Lifshitz hydrodynamics at the first derivative order.\footnote{We would like to thank Shira Chapman for pointing out the possibility of having a non-vanishing $\alpha'$ in the Lifshitz fluid.} They appear in the antisymmetric part of the energy-momentum tensor as 
\begin{equation}
T^{[\mu\nu]}=\pi_A^{[\mu\nu]}=-\alpha u^{[\mu}a^{\nu]}-T(\alpha'+C)u^{[\mu} P^{\nu]\sigma}\partial_\sigma\left(\frac{\mu}{T}\right),
\end{equation} 
where $a^\mu = u^\alpha \partial_\alpha u^\mu $ is the acceleration and $T$, $\mu$ are the temperature and chemical potential. The new transport coefficients also appears in the $U(1)$ current in a different combination $j^\mu\supset  -(\alpha'-C) a^\mu$. The coefficient $C$ is {\em dissipationless} and does not satisfy the Onsager relation.

We will start in section \S 2 by giving a covariant formulation of a free Lifshitz theory and construct an improved energy-momentum tensor that satisfies the trace Ward identity. It is not completely obvious that the identity should be satisfied, since Lifshitz theories have scale but not conformal invariance, so in principle there could be a virial current. We will see that this is actually the case for the free scalar, but nevertheless the energy-momentum tensor can be improved and eliminate the virial term, which is only possible because Lorentz symmetry is explicitly broken.

In \S 3 we will derive the path integral form of the thermal partition function and compute the temperature dependence of one- and two-point functions of the energy-momentum tensor at zero frequency and spatial momentum. We will use these results to write down the ideal hydrodynamic energy-momentum tensor at arbitrary velocities. 

In the last sections we study a generalization of Lifshitz hydrodynamics by adding a conserved current. In \S 4 we follow \cite{Banerjee:2012iz,Jensen:2012jh} to derive the ideal energy-momentum tensor from a generating functional. We find a generalization of Weyl transformations of the metric that reproduces the Ward identity. In \S 5 we use the fluid/gravity correspondence \cite{Rangamani:2008gi} to derive the equations of Lifshitz hydrodynamics at the ideal level. In \S 6 we go beyond the ideal level and consider new possible terms allowed by the breaking of Lorentz symmetry. At first order there is a dissipative term that we already discussed in \cite{Hoyos:2013eza} and for charged hydrodynamics an additional dissipationless term producing a current along the acceleration of the fluid is also possible. We show how to compute the new transport coefficients from the energy-momentum tensor and current correlators using Kubo formulas. We end with a discussion of our results and appendices containing some technical details.

%%%%%%%%%%%%%%%%%%%%%%%%%%%%%%%%%%%%%%%%%%%%%%%%%%%%%%%%%%%
\section{Covariant formulation of Lifshitz theories}
%%%%%%%%%%%%%%%%%%%%%%%%%%%%%%%%%%%%%%%%%%%%%%%%%%%%%%%%%%%

Although we are interested in general interacting theories with Lifshitz scaling, we find  useful to show explicitly some of their properties in a concrete model. For simplicity we will restrict to local theories, $z=2 n$, $n\in \mathbb{Z}$, and in particular to the simplest case $z=2$. A local action for a free scalar can also be written for rational values of the dynamical exponent $z=2n/m$, $n,m\in \mathbb{Z}$,
\begin{equation}
S=\int dt\,d^d x\,\left[\frac{i^m}{2}\phi (\partial_t^m \phi)-\frac{m\kappa}{4n}((\partial_i^2)^{n}\phi)^2 \right].
\end{equation}
For some values of $m$ and $n$, the symmetry is enhanced. For instance, there is a Schr\"odinger symmetry for $m=n=1$  \cite{Mehen:1999nd} and a relativistic conformal symmetry for $m=2$, $n=1$.

The main points that we want to highlight using the free field theory example are:
\begin{itemize}
\item For local Lifshitz theories there is a covariant formulation even though Lorentz invariance is explicitly broken. The Lagrangian is well-defined as a scalar density from which one can derive the energy-momentum tensor. The breaking of Lorentz symmetry is manifested in the {\it asymmetric} properties of the energy-momentum tensor
\begin{equation}
T^{i0} \neq T^{0i}.
\end{equation} 
The Lorentz Ward identity for the energy-momentum tensor
$\vev{T^{ij}}-\vev{T^{ji}}=0$
only holds when the two indices are spatial (assuming rotational invariance).
\item Lifshitz theories in flat space (given $z$ and $ d$) can be defined for arbitrary time-like Killing vectors. These theories are essentially identical since it is possible to do frame transformations to map different Lifshitz theories on each other. This will be important for hydrodynamics because it justifies our formulation where the equation of state is independent of the frame. We will show it explicitly for the $z=2$ theory in section \ref{sec:thermod}.

As an analogy, consider a theory of complex scalars $\phi^a$, $a=1,2$ with a $U(2)$ flavor symmetry. If we add a term to the Lagrangian of the form
\begin{equation}
\Delta \cL\propto \frac{1}{2}(\sigma^3)_{ab}(\phi^a)^\dagger\phi^b,
\end{equation}
the symmetry is explicitly broken to $U(1)\times U(1)$. The change of variables
\begin{equation}
\phi^a\to U^a_{\ b} \phi^b, \  \ U \in SU(2),
\end{equation}
maps the term that breaks the $SU(2)$ symmetry into
\begin{equation}
\Delta \cL' \propto \frac{1}{2}(U^\dagger \sigma^3 U)_{ab}(\phi^a)^\dagger\phi^b.
\end{equation}
Although the couplings in the action look different, the theory with the term $\Delta \cL$ and the theory with $\Delta \cL'$ are equivalent. They are related by a change of variables (all observables are related by a simple $U(2)$ rotation). 

The analogous statement is true for Lifshitz theories, with the subtlety that the theory should be quantized along the direction determined by the time-like Killing vector. Lifshitz theories with different Killing vectors quantized along the same time direction are inequivalent. Typically this breaks the scaling symmetry as well, so we will not consider it here.
\end{itemize}

In order to define the $z=2$ Lifshitz theory covariantly in curved spacetime, we will introduce the vierbein fields $e_\mu^{\ a}$. The action is
\begin{equation}
S=\int d^{d+1} x\, e\left[\frac{1}{2}(\nabla_\parallel\phi)^2-\frac{\kappa}{4}(\nabla_\perp^2\phi)^2 \right]+\cdots,
\end{equation}
where the dots correspond to couplings of the scalar to the curvature, and
\begin{align}
&\nabla_\parallel\phi=t^\mu\nabla_\mu\phi=t^a e_a^{\ \mu}\nabla_\mu\phi,\\
&\nabla_\perp^2\phi=P_t^{\mu\nu}\nabla_\mu\nabla_\nu\phi=P_t^{ab} e_a^{\ \mu}e_b^{\ \nu}\nabla_\mu\nabla_\nu\phi.
\end{align}
Here $t^a=(1,0)$, $P_t^{ab}=\eta^{ab}+t^a t^b$ and $\eta_{ab}t^a t^b=-1$. The flat spacetime action is recovered by substituting $e_\mu^{\ a}=\delta_\mu^{\ a}$.
 
From the action we can extract the Lagrangian density
\begin{equation}
\cL=\frac{1}{2}(\nabla_\parallel\phi)^2-\frac{\kappa}{4}(\nabla_\perp^2\phi)^2.
\end{equation}
The Lagrangian density is explicitly invariant under the coordinate transformations:
\begin{equation}
\nabla_\mu\phi\to \Lambda_\mu^\nu \nabla_\nu\phi, \ \ e_\mu^{\ a}\to e_\nu^{\ a}\left(\Lambda^{-1}\right)^\nu_\mu.
\end{equation}
In flat space we can do a coordinate transformation combined with a frame transformation that leaves the vierbein (and hence the metric) invariant
\begin{equation}
e_\mu^{\ a}\to \Lambda^a_{b} e_\nu^{\ b} \left(\Lambda^{-1}\right)^{ \nu}_\mu= e_\mu^{\ a}=\delta_\mu^a.
\end{equation}
Note that, contrary to a relativistic theory, the Lagrangian density is in general not invariant under this transformation
\begin{align}
\cL &=(\Lambda^T)_{c}^{ a} t^c t^d (\Lambda)_d^b e_a^\mu e_b^\nu \nabla_\mu\phi\nabla_\nu\phi \nonumber \\
& +(\Lambda^T)_{a'}^a (\Lambda^T)_{c'}^c P_t^{a'b'}P_t^{c'd'}(\Lambda)_{b'}^b (\Lambda)_{d'}^d e_a^\mu e_b^\nu e_c^\lambda e_d^\rho\nabla_\mu\nabla_\nu\phi\nabla_\lambda\nabla_\rho\phi.
\end{align}
If we do the combined coordinate plus frame transformation that leaves the background vierbeins invariant, a frame transformation rotates the unit time vector $t^a$ into 
\begin{equation}
u^a=(\Lambda^T)_c^a t^c.
\end{equation}
Therefore, Lifshitz theories defined with different vectors $u^a$ are equivalent.

%%%%%%%%%%%%%%%%%%%%%%%%%%%%%%%%%%%%%%%%%%%%%%%%%%%%%%%%%%%
\subsection{Canonical energy-momentum tensor}
%%%%%%%%%%%%%%%%%%%%%%%%%%%%%%%%%%%%%%%%%%%%%%%%%%%%%%%%%%%

Having determined the action in curved spacetime, we can extract the energy-momentum tensor from the variation of the action with respect to the vierbein
\begin{equation}
T^\alpha_{\ c}=-\frac{1}{e}\frac{\delta S}{\delta e_\alpha^{ \ c}}.
\end{equation}
In contrast with the usual definition using the variation with respect to the metric, in principle the energy-momentum tensor is not necessarily symmetric in its two indices. The Lifshitz scaling symmetry should translate into a `trace' Ward identity for the energy-momentum tensor, similar to the condition $T^\mu_{\ \mu}=0$ in a conformal field theory. 

In many cases the trace identity is not satisfied by the energy-momentum tensor derived from the na\"{\i}ve  extension of the action to curved space. It is necessary to add improvement terms that do not affect to the conservation equations. We will find, in the free scalar example, that a term proportional to the divergence of a virial current $V^\mu$ remains, while most of the contributions can be improved by adding couplings of the scalar field to the curvature. For a general Lifshitz theory, we thus expect
\begin{equation} \label{TrceGeneral2} 
zt_\alpha t^a T^\alpha_{\ a}-P^a_{t\,\alpha} T^{\alpha}_{\ a} = \partial_\alpha V^\alpha=\partial_\parallel V^\parallel+P_t^{\alpha\beta}\partial_\alpha V_\beta.
\end{equation}
Where $V^\parallel=-t^\alpha V_\alpha$. 

The virial current cannot be improved unless it is of the form $V_\mu=\partial_\mu X$ in a relativistic theory  (c.f.~\cite{Polchinski:1987dy}). This is not the case in the example we study. However, because Lifshitz is not Lorentz invariant, there is still room to add further improvement terms that fix the trace Ward identity. Both terms in (\ref{TrceGeneral2}) depend on the virial current and can be removed by adding the improvement terms
\begin{equation}\label{improveq}
\tilde T_{\ c}^\alpha=T_{\ c}^\alpha-\frac{1}{d}\left(t^\alpha  \partial_c V^\parallel -\delta^\alpha_{c}\partial_\parallel V^\parallel\right)-\frac{1}{z-1+d}\left(P_t^{\alpha\beta}\partial_cV_\beta-\delta^\alpha_cP_t^{\sigma\rho}\partial_\sigma V_\rho\right).
\end{equation}
Note that these terms give asymmetric contributions to the energy-momentum tensor (even if $z=1$). Thus we cannot improve the energy-momentum tensor and preserve Lorentz invariance in general. We will later use the Ward identity to constrain the form of the energy-momentum tensor in ideal hydrodynamics. 

In the following we will derive the energy-momentum tensor for the $z=2$ Lifshitz theory with an arbitrary time-like vector. We will show that it is possible to improve the energy-momentum tensor to satisfy a trace Ward identity, which depends on the time-like vector. There is a non-zero virial current. It is possible to improve the virial current without affecting the Ward identity. 

The canonical energy-momentum tensor in flat space is given by 
\begin{align}
\notag T_{\rm can\ c}^\alpha &=\partial_\parallel \phi t^\alpha \partial_c\phi-\kappa P_t^{\alpha\sigma}\partial_c\partial_\sigma\phi \partial_\perp^2\phi-\delta^\alpha_c \cL\\
&+\frac{\kappa}{2}\left(P_{t\, c}^{\ \beta}\partial_\beta(\partial^\alpha\phi  \partial_\perp^2\phi)- P_{t\, c}^{\ \alpha}\partial_\beta(\partial^\beta\phi  \partial_\perp^2\phi) +P_t^{\alpha\beta}\partial_\beta(\partial_c\phi\partial_\perp^2\phi)\right).
\end{align}
We use the formulas in Appendix~\ref{sec:vierbeins} to compute the variation of the action and set $e_\mu^{\ a}=\delta_\mu^a$ at the end of the calculation to get the result.  
One can show explicitly that it is conserved 
\begin{equation}
\partial_\alpha T^\alpha_{\rm can\ c}=\partial_c\phi\left[\partial_\parallel^2\phi+\frac{\kappa}{2}(\partial_\perp^2)^2\phi \right]=0.
\end{equation}

To find the trace, we compute the projection on $t$ and the transverse directions of the energy-momentum tensor 
\begin{align}
&T^\alpha_{\rm can\ c} t^c t_\alpha=-\frac{1}{2}(\partial_\parallel\phi)^2-\frac{\kappa}{4}(\partial_\perp^2\phi)^2,\\
&T^\alpha_{\rm can\ c} P_{t\,\alpha}^{\ c}=-\frac{d}{2}(\partial_\parallel\phi)^2+\kappa\frac{d}{4}(\partial_\perp^2\phi)^2+\kappa\left[ P_t^{\alpha\beta}\partial_\alpha\phi\partial_\beta\partial_\perp^2\phi-\frac{d}{2}\partial_\beta(\partial^\beta\phi\partial_\perp^2\phi)\right].
\end{align}
One can use that $\partial^2=\partial_\perp^2-\partial_\parallel^2$ and $\partial_\alpha X\partial^\alpha Y=P_t^{\alpha\beta}\partial_\alpha X\partial_\beta Y-\partial_\parallel X\partial_\parallel Y$ to rewrite the last expression as
\begin{align}
T^\alpha_{\rm can\ c} P_{t\,\alpha}^{\ c} &=-\frac{d}{2}(\partial_\parallel\phi)^2-\kappa\frac{d}{4}(\partial_\perp^2\phi)^2-\kappa\left(\frac{d}{2}-1\right)P_t^{\alpha\beta}\partial_\alpha\phi \partial_\beta \partial_\perp^2\phi \nonumber \\
&+\kappa\frac{d}{2}(\partial_\parallel^2 \phi \partial_\perp^2\phi+\partial_\parallel \phi \partial_\parallel \partial_\perp^2\phi).
\end{align}

The Lifshitz trace Ward identity is 
\begin{equation}
{\rm Tr}=2 T^\alpha_{\ c} t^c t_\alpha-T^\alpha_{\ c} P_{t\,\alpha}^{\ c}=0.
\end{equation}
For the canonical tensor, we have 
\begin{equation}
{\rm Tr}_{\rm can}=\left(\frac{d}{2}-1\right)\left[ (\partial_\parallel\phi)^2+\frac{\kappa}{2}(\partial_\perp^2\phi)^2+\kappa P_t^{\alpha\beta}\partial_\alpha\phi \partial_\beta \partial_\perp^2\phi\right]-\kappa\frac{d}{2}(\partial_\parallel^2 \phi \partial_\perp^2\phi+\partial_\parallel \phi \partial_\parallel \partial_\perp^2\phi).
\end{equation}
In the following we add improvement terms to check whether we can make the trace to vanish.

%%%%%%%%%%%%%%%%%%%%%%%%%%%%%%%%%%%%%%%%%%%%%%%%%%%%%%%%%%%
\subsection{Improvement terms and trace Ward identity}
%%%%%%%%%%%%%%%%%%%%%%%%%%%%%%%%%%%%%%%%%%%%%%%%%%%%%%%%%%%

We can add the following terms to the action
\begin{equation}
S_R=-\int d^{d+1} x e\left[c_1 R_{\mu\nu} t^\mu t^\nu X +c_2 R_{\mu\nu} P_t^{\mu\nu} Y  \right].
\end{equation}
The variation of the action around flat space gives
\begin{align}
T^\alpha_{R\ c} &=c_1  (\delta_c^{\ \alpha}\partial_\parallel^2-t_c\partial^\alpha \partial_\parallel+ t_c t^\alpha\partial^2-t^\alpha \partial_c \partial_\parallel) X \nonumber \\
&+c_2 (\delta_c^{\ \alpha}\partial_\perp^2-P_{t\,c}^{\ \beta}\partial^\alpha \partial_\beta+ P_{t\,c}^{\ \alpha}\partial^2-P_t^{\alpha\beta} \partial_c\partial_\beta) Y.
\end{align}
This contribution does not affect to the conservation of the energy-momentum tensor $\partial_\alpha T^\alpha_{R\ c}=0.$ 
For the contributions to the trace, we compute the projections of $T^\alpha_{R\ c}$
\begin{align}
T^\alpha_{R\ c} t_\alpha t^c &=c_1\partial_\perp^2 X-c_2 \partial_\perp^2 Y, \\ 
T^\alpha_{R\ c} P_{t\,\alpha}^{\ c} &= d c_1\partial_\parallel^2 X+c_2(2(d-1) \partial_\perp^2 -d\partial_\parallel^2 )Y.
\end{align}
Then, the total contribution to the trace is
\begin{equation}  \label{TraceImprove}
{\rm Tr}_R=c_1(2\partial_\perp^2-d \partial_\parallel^2)X+c_2(d\partial_\parallel^2-2d \partial_\perp^2)Y .
\end{equation}

If instead we use the combinations defined in (\ref{Rcombinations}) $
R_\parallel =-\frac{1}{d-1}\left(R_{\mu\nu} t^\mu t^\nu+\frac{1}{d}R_{\mu\nu}P_t^{\mu\nu}\right),$ 
$R_\perp =-\frac{1}{2(d-1)}\left(R_{\mu\nu} t^\mu t^\nu +R_{\mu\nu}P_t^{\mu\nu}\right)$ 
in the following action 
\begin{equation}
S_R=-\int d^{d+1} x e\left[c_\parallel R_\parallel X +c_\perp R_\perp	 Y  \right],
\end{equation}
then we can read off the trace from (\ref{TraceImprove}). We get
${\rm Tr}_R=c_\parallel \partial_\parallel^2 X+c_\perp \partial_\perp^2Y.$
Then, with the choice 
\begin{align}   \label{ImproveCoeff}
c_\parallel = \frac{c_\perp}{\kappa} =-\frac{1}{2} \left(\frac{d}{2}-1\right) , \quad 
X=\phi^2 , \quad  Y=\phi \partial_\perp^2 \phi \;,   
\end{align}
the trace becomes
\begin{align}
\notag {\rm Tr}_{\rm can}+{\rm Tr}_R &=-\left(\frac{d}{2}-1\right)\phi\left[\partial_\parallel^2\phi+\frac{\kappa}{2}(\partial_\perp^2)^2 \phi \right]-\kappa\frac{d}{2}(\partial_\parallel^2 \phi \partial_\perp^2\phi+\partial_\parallel \phi \partial_\parallel \partial_\perp^2\phi)\\
&=-\kappa\frac{d}{2}(\partial_\parallel^2 \phi \partial_\perp^2\phi+\partial_\parallel \phi \partial_\parallel \partial_\perp^2\phi)=-\kappa\frac{d}{2}\partial_\parallel(\partial_\parallel \phi \partial_\perp^2\phi) ,
\end{align}
where we have used the equations of motion, and 
\begin{align}
\frac{1}{2}\partial_\parallel^2 \phi^2 &=(\partial_\parallel \phi)^2+\phi \partial_\parallel^2\phi, \ \\  \partial_\perp^2(\phi \partial_\perp^2\phi) &=(\partial_\perp^2\phi)^2+2 P_t^{\mu\nu}\partial_\mu \phi \partial_\nu (\partial_\perp^2)^2 \phi+\phi (\partial_\perp^2)^2 \phi. 
\end{align}

In principle we would like to find a coupling to the curvature that can cancel out the leftover part. However, it is not strictly necessary. For the energy-momentum tensor $T^\alpha_{\ c}=T^\alpha_{{\rm can}\ c}+T^\alpha_{R\ c}$, we can define a conserved dilatation current 
\begin{equation}
D^\alpha=2 T^\alpha_{\ c}t^c t_\mu x^\mu-T^\alpha_{\ c} P_{t\ \mu}^c x^\mu-t^\alpha V,
\end{equation}
where $t^\alpha V$ is analogous to the virial current 
\begin{equation}
V=-\kappa\frac{d}{2}\partial_\parallel \phi \partial_\perp^2\phi.
\end{equation}
Then, one can show 
\begin{equation}
\partial_\alpha D^\alpha= 2  T^\alpha_{\ c}t^c t_\alpha-T^\alpha_{\ c} P_{t\ \alpha}^c -\partial_\parallel V=0.
\end{equation} 

From now, we take $d=2$. Then $T^\alpha_{R\, c}=0$, which can be checked easily from (\ref{ImproveCoeff}). 
We define the conserved energy current as
\begin{equation}
\cE^\alpha=-T^\alpha_{\ c}t^c,
\end{equation}
where $ T^\alpha_{\ c} = T^\alpha_{\text{can} \ c}$ 
or $ T^\alpha_{\ c} = T^\alpha_{\text{can} \ c} + T^\alpha_{R \ c}$ for $d=2$. 
Concretely, we take 
\begin{equation}\label{energy}
\cE^\alpha=-t^\alpha\left[\frac{1}{2}(\partial_\parallel\phi)^2+\frac{\kappa}{4}(\partial_\perp^2\phi)^2\right]+\frac{\kappa}{2}P^{\alpha\beta}_{t}\left[\partial_\parallel\partial_\beta\phi\partial_\perp^2\phi -\partial_\parallel\phi\partial_\beta\partial_\perp^2\phi \right].
\end{equation}
The projections, longitudinal and transverse to $t$, are
\begin{align}
\label{t00} t_\alpha \cE^\alpha &=\frac{1}{2}(\partial_\parallel\phi)^2+\frac{\kappa}{4}(\partial_\perp^2\phi)^2,\\
\label{ti0} P^\alpha_{t\,\ \beta}\cE^\beta &=\frac{\kappa}{2}P^{\alpha\beta}_{t}\left[\partial_\parallel\partial_\beta\phi\partial_\perp^2\phi -\partial_\parallel\phi\partial_\beta\partial_\perp^2\phi \right].
\end{align}

The conserved momentum current is defined in a similar way
\begin{equation}
\cP^\alpha_{\ c}=-T^\alpha_{\ b}P^b_{t\ c}.
\end{equation}
We have
\begin{align}\label{momentum}
\notag \cP^\alpha_{\ c} & =-t^\alpha P^b_{t\ c}\left[\partial_\parallel \phi\partial_b\phi+\frac{1}{2}\partial_b V \right]-\frac{\kappa}{2}P^{\alpha\beta}_{t}P^{b}_{t\ c} \left[\partial_\beta\phi\partial_b\partial_\perp^2\phi+\partial_b\phi\partial_\beta\partial_\perp^2\phi \right]\\ &+P^{\alpha}_{t\ c}\left[\frac{1}{2}(\partial_\parallel\phi)^2 +\frac{\kappa}{4}(\partial_\perp^2\phi)^2+ \frac{\kappa}{2}P_t^{ab}\partial_a\phi\partial_b\partial_\perp^2\phi+ \frac{1}{2}\partial_\parallel V\right].
\end{align}
Similarly, the projections are
\begin{align}
\label{t0i} t_\alpha \cP^\alpha_{\ c} &=P^b_{t\,c}\left[\partial_\parallel \phi\partial_b\phi+\frac{1}{2}\partial_b V\right],\\
\notag P^\alpha_{t\,\ \beta} \cP^\beta_{\ c} &=-\frac{\kappa}{2}P^{\alpha\beta}_{t}P^{b}_{t\ c} \left[\partial_\beta\phi\partial_b\partial_\perp^2\phi+\partial_b\phi\partial_\beta\partial_\perp^2\phi \right]\\ \label{tij}
&+P^{\alpha}_{t\ c}\left[\frac{1}{2}(\partial_\parallel\phi)^2 +\frac{\kappa}{4}(\partial_\perp^2\phi)^2+ \frac{\kappa}{2}P_t^{ab}\partial_a\phi\partial_b\partial_\perp^2\phi+\frac{1}{2}\partial_\parallel V\right].
\end{align}
Note that all the terms in each expression have the same scaling except for the ones depending on the virial terms. 

We can get rid of the virial terms by redefining the momentum current to
\begin{equation}\label{improvedP}
\tilde\cP^\alpha_{\ c}=\cP^\alpha_{\ c}-\frac{1}{2}\left(t^\alpha P^b_{t\,c} \partial_b V -P^\alpha_{t\,c}\partial_\parallel V\right).
\end{equation}
One can check the new momentum current is still conserved $\partial_\alpha\tilde \cP^\alpha_{\ c}=0$. Note that, with this definition, all the components of the momentum current have the same scaling, and the Ward identity is satisfied
\begin{equation}
2t_\alpha\cE^\alpha-P^c_{t\, \alpha}\tilde\cP^\alpha_{\ c}=0.
\end{equation}
In fact we can see the new term in the definition of the momentum as an improvement of the energy-momentum tensor
\begin{equation}
\tilde T_{\ c}^\alpha=T_{\ c}^\alpha-\frac{1}{2}\left(t^\alpha  \partial_c V -\delta^\alpha_{c}\partial_\parallel V\right).
\end{equation}
Note that $V$ is \emph{not} a total derivative, thus the improvement terms cannot come from curvature terms added to the action. One can check that the improvement terms do not contribute to the energy current. With the improved tensor, the dilatation current will take the usual form $\tilde D^\alpha=2\tilde T_{\ c}^\alpha t^c t_\mu x^\mu - \tilde T_{\ c}^\alpha P^c_{t\ \mu}x^\mu$ up to a total derivative. The general form of the improvement term is given in \eqref{improveq}. Assuming that $V^a$ is independent of the vierbeins, it can be derived from an added term to the action of the form
\begin{equation}
S_{\rm improv}= \frac{1}{d}\int d^dx\,|e|e_a^{\ \mu}\left(t^a t_b-\frac{d}{z-1+d}P^a_{t\ b} \right)\partial_\mu V^b.
\end{equation}

%%%%%%%%%%%%%%%%%%%%%%%%%%%%%%%%%%%%%%%%%%%%%%%%%%%%%%%%%%%
\section{Free Lifshitz theories at finite temperature}\label{sec:thermod}
%%%%%%%%%%%%%%%%%%%%%%%%%%%%%%%%%%%%%%%%%%%%%%%%%%%%%%%%%%%

At finite temperature the energy-momentum tensor acquires a non-zero expectation value, and two-point functions receive temperature-dependent contributions at zero momentum. We will use the $z=2$ free scalar example to compute them explicitly and then generalize the formulas for other values of $z$. First we will find the scaling properties of one- and two- point functions in terms of the temperature. Then we check the trace Ward identity for a Lifshitz theory defined with an arbitrary time-like vector. 

As we will see, the energy-momentum tensor takes the same form as the ideal energy-momentum tensor in hydrodynamics. 
Furthermore, the Ward identity fixes the equation of state in an arbitrary frame
\begin{equation}
z t_\mu t^\alpha T^\mu_{\ \alpha}-P^\mu_{t\ \alpha} T^{\mu}_{\ \alpha}=-z\varepsilon_0+ dp_0=0,
\end{equation}
where the energy density and pressure are defined from the one-point functions when $t^\alpha=(1,\mathbf{0})$
\begin{equation}
\vev{T_{00}}=\varepsilon_0, \ \ \vev{T_{ij}}=p_0\delta_{ij}.
\end{equation}

After the analysis in flat spacetime we will use the generating functional approach of  \cite{Banerjee:2012iz,Jensen:2012jh} to find equilibrium configurations in a curved spacetime. We find that the spatial components of the velocities depend, to linear order, on the mixed components of the vierbein 
\begin{equation}
u^i \simeq  C_1 e_0^{\ i}+C_2 e_j^{\ 0}\delta^{ij},
\end{equation}
where the coefficients $C_1$ and $C_2$ depend on the two-point functions of the $T^0_{\ i}$ and $T^i_{\ 0}$ components of the energy-momentum tensor. 

In a theory with Lifshitz scaling the energy-momentum correlators depend on a power of the temperature, but in contrast to relativistic theories, the power is different for different correlators.
One could use this result to distinguish between Lifshitz theories and (non-conformal) relativistic theories with the same equation of state. Another outcome of our analysis is that no additional terms are necessary in curved spacetime at the  ideal level. This provides a microscopic justification of our proposal for Lifshitz hydrodynamics that we present in the next sections. In particular, no new dissipative terms are present in ideal hydrodynamics, even though probes may experience a drag force independent of the temperature as in \cite{Kiritsis:2012ta}.

\subsection{General temperature dependence of one- and two- point functions}

We can use scaling arguments to determine the temperature dependence of the energy-momentum correlators. We have  computed, in Appendix~\ref{sec:thermal}, the one- and two- point functions of a free scalar theory at finite temperature. The one-point functions of the energy-momentum tensor for $t^\alpha=(1,\vec{0})$ can be computed using \eqref{improvedP} and \eqref{t00}, \eqref{t0i}, \eqref{ti0} and \eqref{tij}. 

The scaling dimension of the temperature is the same as the scaling of time derivatives $\partial_\parallel \sim \partial_\perp^2\sim T$. Therefore, the scaling with the temperature of the components of the energy-momentum tensor is 
\begin{equation}
\vev{T^{00}}\sim \vev{T^{ij}}\sim T^2, \quad \vev{T^{0i}}\sim T^{\frac{3}{2}}, \quad  \vev{T^{i0}}\sim T^{\frac{5}{2}}.
\end{equation}
We have checked that the Ward identity is satisfied
\begin{equation}
\vev{T^{ij}}=\delta^{ij}\vev{T^{00}}.
\end{equation}
For general dimensions and values of the dynamical exponent 
\begin{equation}
\vev{T^{ij}}=\frac{z}{d}\delta_{ij}\vev{T^{00}}\sim T^{\frac{d+z}{z}}, \quad \vev{T^{0i}}\sim T^{\frac{d+1}{z}}, \quad  \vev{T^{i0}}\sim T^{\frac{d-1+2z}{z}}.
\end{equation}
The expectation values of $\vev{T^{i0}}$ and $\vev{T^{0i}}$ are zero in flat spacetime.

\paragraph{Two-point functions}

We are interested in computing the zero frequency and momentum correlators 
\begin{equation}
C_{\mu\nu,\alpha\beta}=\lim_{\omega\to 0}\lim_{k\to 0}\vev{T_{\mu\nu} T_{\alpha\beta}}(\omega,k).
\end{equation}
We are interested, in particular, in the $T_{0i}$ and $T_{i0}$ values for $d=2$ spatial dimensions. The scaling with the temperature is 
\begin{align}
C_{0i,0j} \sim  T , \quad C_{0i,j0} =C_{i0,0j} \sim T^2, \quad C_{i0,j0} \sim T^3.
\end{align}
The generalization for arbitrary $z$ and $d$ is
\begin{align}
C_{0i,0j} \sim  T^{\frac{d+2-z}{z}} , \quad
C_{0i,j0}=C_{i0,0j} \sim T^{\frac{d+z}{z}}, \quad
C_{i0,j0} \sim T^{\frac{d+3z-2}{z}}.
\end{align}

We can compare these results with the scaling obtained in \cite{Korovin:2013bua} in the zero temperature theory using general field theory arguments. To do so, we convert our results to those of the coordinate space, using 
\begin{align}
\langle \mathcal O (x) \mathcal O(0) \rangle \sim x^{-2\Delta}  \quad \Leftrightarrow \quad
\langle \mathcal O (k) \mathcal O(-k) \rangle_{k \to 0} \sim T^{\frac{2\Delta -d-z}{z}} \;. 
\end{align}
Then we get 
\begin{gather}
\vev{T_{0i} T_{0j}}(t,x) \sim x^{-2-2d}   \;, \quad
\vev{T_{0i} T_{j0}}(t,x) \sim x^{-2d-2z}   \;, \\
\vev{T_{i0} T_{j0}}(t,x) \sim x^{2-2d-4z}  \;.
\end{gather}
In \cite{Korovin:2013bua}, the correlation functions for the modified stress tensor $J_{\mu\nu} = T_{\mu\nu} + A^B_{\mu} J_{\nu} $, with massive gauge field $A^B_{\mu}$ and associated current $ J_{\nu} $, are evaluated for $d=1$ and $z=1 + \epsilon^2$, $ \epsilon \ll 1$ 
\begin{gather}
\vev{J_{tx}(t,x) J_{tx}(0)} \sim x^{-4z}   \;, \quad 
\vev{J_{xt}(t,x) J_{tx}(0)} \sim x^{-2-2z}   \;, \\
\vev{J_{xt}(t,x) J_{xt}(0)} \sim x^{-4}  \;.
\end{gather}
These two results are consistent if we identify $T_{0i} \sim J_{xt} $ and $T_{i0} \sim J_{tx} $.
Our notation and the corresponding scaling are consistent with the conservation equation $\partial_\mu T^{\mu\nu} = 0 $

\subsection{Partition function and energy-momentum tensor in a general frame}

The finite temperature partition function for a field theory is defined as
\begin{equation}
Z_{QFT}[\beta]={\rm Tr}\,\left( e^{-\beta H}\right),
\end{equation}
where $H=T^0_{\ 0}$ is the Hamiltonian. For a Lifshitz theory this definition is not frame-independent, but we can define a frame-independent partition function which coincides with the usual definition when $t^a=(1,\vec{0})$:
\begin{equation}
Z_{\rm Lif}[\beta]={\rm Tr}\,\left( e^{-\beta\int d^dx_\perp\,  t_\alpha \cE^\alpha}\right).
\end{equation}
Where
\begin{equation}
\int d^dx_\perp=\int d^{d+1}x\, \delta(t_\alpha x^\alpha).
\end{equation}
Note that $t^\alpha$ is a time-like Killing vector of spacetime and that we can identify $x^\parallel=t_\alpha x^\alpha$ with the direction where time evolution takes place. The conjugate momentum to $\phi$ is then $\pi=\partial_\parallel\phi$. 

We can follow the usual steps to rewrite the partition as an Euclidean path integral. First we introduce the identity 
\begin{equation}
1=\int d\phi |\phi\rangle\langle\phi|
\end{equation}
$N$-times inside the trace, and use that
\begin{equation}
\vev{\phi_{i+1}|\pi_{i}}=\exp\left(i\int d^d x_\perp\,\pi_{i}\phi_{i+1}\right).
\end{equation}
The partition function takes the form
\begin{align}
& Z_{\rm Lif}[\beta]=\lim_{N\to\infty}\cN\int \prod_{i=1}^N d\pi_i d\phi_i \exp\left(-\frac{\beta}{N}\sum_{i=0}^N \int d^dx_\perp\,\left[t_\alpha\cE^\alpha_i -i\pi_i\frac{(\phi_{i+1}-\phi_i)}{\beta/N}\right] \right).
\end{align}
It is understood that $\phi_0=\phi_N$, and $\cN$ is a suitable normalization. 

For the $z=2$ free scalar case, we found
\begin{equation}
t_\alpha \cE^\alpha=\frac{1}{2}\pi^2+\frac{\kappa}{4}(\partial_\perp^2\phi)^2.
\end{equation}
The integrals over momenta are Gaussian, so we can make them explicitly. This leads to
\begin{align}
& Z_{\rm Lif}[\beta]=\lim_{N\to\infty}\cN\int \prod_{i=1}^N  d\phi_i \exp\left(-\frac{\beta}{N}\sum_{i=0}^N \int\,d^dx_\perp\,\left[\frac{1}{2}\left(\frac{(\phi_{i+1}-\phi_i)}{\beta/N}\right)^2+\frac{\kappa}{4}(\partial_\perp^2\phi_i)^2\right] \right).
\end{align}
Taking the continuum limit, we are left with the Euclidean path integral
\begin{align}
& Z_{\rm Lif}[\beta]=\int \cD\phi(\tau,x) \exp\left(-\int_0^\beta d\tau\,\int\,d^dx_\perp\,\left[\frac{1}{2}\left(\partial_\tau\phi\right)^2+\frac{\kappa}{4}(\partial_\perp^2\phi)^2\right] \right).
\end{align}
The field $\phi$ is periodic in the $\tau$ direction as usual. This shows that in order to compute thermal correlators we can simply change $\partial_\parallel$ derivatives into $\partial_\tau$ derivatives as we do in usual QFT calculations with time derivatives.

In order to compute the expectation values of the energy and momentum currents, we should use the expressions \eqref{energy}, \eqref{momentum} and \eqref{improvedP}. For the energy we get\footnote{There is temperature-independent divergent contribution that needs to be subtracted in order to get a finite energy density. The same applies to other expressions involving integrals over momentum.}
\begin{align}
&\vev{\cE^\alpha}=\frac{t^\alpha}{2\beta}\sum_n \int \frac{d^2 q_\perp}{(2\pi)^2}\frac{\omega_n^2+\frac{\kappa}{2}(q_\perp^2)^2}{\omega_n^2-\frac{\kappa}{2}(q_\perp^2)^2}= -\varepsilon t^\alpha.
\end{align}
Where we have used that terms odd in the momentum vanish. For the momentum we will have
\begin{align}
&\vev{\cP^\alpha_c}=\frac{P_{t\ c}^\alpha}{2\beta}\sum_n \int \frac{d^2 q_\perp}{(2\pi)^2}\frac{\omega_n^2+\frac{\kappa}{2}(q_\perp^2)^2}{\omega_n^2-\frac{\kappa}{2}(q_\perp^2)^2}= p P_{t\ c}^\alpha=\varepsilon P_{t\ c}^\alpha.
\end{align}
In this case we have used that
\begin{equation}
\int d^2 q_\perp \, q_\perp^\alpha q^\beta_\perp = \frac{1}{2}P_t^{\alpha\beta}\int d^2 q_\perp  q_\perp^2.
\end{equation}
This shows that the expectation value of the (improved) energy-momentum tensor at finite temperature in an arbitrary Lifshitz theory defined by $t^\alpha$ has the form
\begin{equation}
\vev{T^\alpha_{\ c}}=\varepsilon t^\alpha t_c+p P^\alpha_{t\ c},
\end{equation}
and the equation of state is independent of $t^\alpha$
\begin{equation}
\varepsilon = p.
\end{equation}
Clearly, this can be generalized for other values of $z$ and $d$. 

We identify the Lifshitz theory defined with $t^a=(1,\vec{0})$ as the one that determines the properties of the fluid at rest. When the fluid is moving with a velocity $u^\mu$ with respect to an observer, the (ideal) energy-momentum tensor measured by the observer will be
\begin{equation}
\vev{T^\alpha_{\ c}}=\varepsilon u^\alpha u_c+p P^\alpha_{u\ c}.
\end{equation}
This is just the same form as the energy-momentum tensor in a relativistic theory in flat spacetime. As an operator $T^\alpha_{\ c}$ is not symmetric, so differences can appear when going beyond the ideal level or in curved spacetimes.

\subsection{Equilibrium hydrodynamics in curved space}

In curved spacetime the expectation value of the energy-momentum is modified by the presence of a background metric/vierbein. If the deviation from flat space is small, we can use linear response theory to find the change to leading order. For the relativistic theory
\begin{equation}
\delta \vev{T^{\mu\nu}(x)}= \int d^{d+1} x'\, G_R^{\mu\nu\alpha\beta}(x,x')\delta g_{\alpha\beta}(x'),
\end{equation}
where $\delta g_{\mu\nu}=g_{\mu\nu}-\eta_{\mu\nu}$ and $G_R$ is the retarded correlator
\begin{equation}
G_R^{\mu\nu\alpha\beta}(x,x')=i\Theta(t-t')\vev{[T^{\mu\nu}(x), T^{\alpha\beta}(x')]}.
\end{equation}
Since in flat space there is translation invariance, the retarded correlator actually depends only on the difference $x-x'$. One can do a Fourier transform (omitting spacetime indices)
\begin{equation}
G_R(\omega,\mathbf{q})=\int dt d^d \mathbf{x} e^{-i\omega(t-t')+i \mathbf{q}\cdot (\mathbf{x}-\mathbf{x'})}G_R(t-t',\mathbf{x}-\mathbf{x'}).
\end{equation}
The zero frequency limit coincides with the Euclidean propagator
\begin{equation}
G_R(\omega=0,\mathbf{q})=G_E(0,\mathbf{q}).
\end{equation}

If we use the hydrodynamic constitutive relations
\begin{equation}
\vev{T^{\mu\nu}}=(\varepsilon+p)u^\mu u^\nu+pg^{\mu\nu},
\end{equation}
and solve the ideal hydrodynamic equations to leading order in $\delta g_{\mu\nu}$
\begin{equation}
\nabla_\mu T^{\mu\nu}=0,
\end{equation}
the prediction from hydrodynamics in a relativistic theory is that at zero frequency and momentum c.f.~\cite{Banerjee:2012iz}
\begin{equation}
\vev{T_{i0} T_{j0}}(\omega=0,\mathbf{q}=0)=\vev{T_{i0} T_{0j}}(\omega=0,\mathbf{q}=0)=\frac{1}{2}\vev{T_{00}}\delta_{ij}.
\end{equation}
The same relation will hold if one derives the hydrodynamic energy-momentum tensor from a generating functional as is done in \cite{Banerjee:2012iz} as well. We have  computed in Appendix~\ref{sec:thermal} the one- and two- point functions of a free scalar theory at finite temperature. One can easily check that the relation above is satisfied in a relativistic theory, but it does not hold in a general theory with Lifshitz scaling. 

In order to see how the difference is manifested in hydrodynamics, we will follow the approach of \cite{Banerjee:2012iz,Jensen:2012jh} to derive the ideal hydrodynamic energy-momentum tensor from the generating functional of a theory in curved space at thermal equilibrium. We will introduce a `Killing vector field' that depends on the vierbeins. The generating functional is
\begin{equation}
W=\int d^d x\, |e| p(T),
\end{equation}
where
\begin{equation}
T=\frac{T_0}{\sqrt{-(eV)\cdot(eV)}}=\frac{T_0}{\sqrt{-(eV)^2}}.
\end{equation}
We are using the notation
\begin{equation}
(eV)^a=e_\mu^a V^\mu, \ \ (eV)\cdot (eV)=\eta_{ab}(eV)^a (eV)^b.
\end{equation}
The variation of the temperature with respect to the vierbein is
\begin{equation}
\frac{\delta T}{\delta e_\mu^{\ a}}=T F^\mu_{\ a}=T\left[u^\mu (eu)_{a} +\frac{1}{\sqrt{-(eV)^2}}\left(e\frac{\delta V}{\delta e_\mu^{\ a}}\right)\cdot (eu) \right].
\end{equation}
We have defined
\begin{equation}
u^\mu=\frac{V^\mu}{\sqrt{-(eV)^2}}.
\end{equation}
Note that
\begin{equation}\label{eqF}
F^\mu_{\ a}\equiv \sqrt{-(eV)^2}\frac{\delta}{\delta e_\mu^{\ a}} \frac{1}{\sqrt{-(eV)^2}}=-\frac{1}{2}\frac{\delta}{\delta e_\mu^{\ a}} \log\left(-(eV)^2\right).
\end{equation}

The expectation value of the energy-momentum tensor is the variation of the generating functional with respect to the vierbein:
\begin{equation}
\vev{\cT^\mu_{\ a}}=\frac{1}{|e|}\frac{\delta W}{\delta e^{\ a}_\mu}=T\frac{\partial p}{\partial T} F^\mu_{\ a} + e^\mu_{\ a} p.
\end{equation}
In flat space $e_\mu^{\ a}=\delta_\mu^{\ a}$ we recover the usual energy-momentum tensor
\begin{equation}
\vev{\cT^\mu_{\ a}}=T\frac{\partial p}{\partial T} u^\mu (\delta u)_a + p \delta^\mu_{\ a},
\end{equation}
provided
\begin{equation}
\left(\delta\frac{\delta V}{\delta e^{\ a}_\mu}\right)\cdot (\delta  u)=\eta_{\alpha\beta}\frac{\delta V^\alpha}{\delta e_\mu^{\ a} }u^\beta =0.
\end{equation}

In order to compute the two-point function we take the second variation of the generating functional.  
\begin{align}
&2 \vev{\cT^\mu_{\ a} \cT^\nu_{\ b}}=\frac{1}{|e|}\frac{\delta^2 W}{\delta e_\mu^{\ a} \delta e_\nu^{\ b}}\\
\notag &=(e^\mu_{\ a} e^\nu_{\ b}-e^\mu_{\ b} e^\nu_{\ a} )p+T\frac{\partial p}{\partial T}(e^\mu_{\ a} F^\nu_{\ b}+e^\nu_{\ b} F^\mu_{\ a})+T\frac{\partial}{\partial T}\left(T\frac{\partial p}{\partial T}\right)F^\mu_{\ a} F^\nu_{\ b}+T\frac{\partial p}{\partial T}\frac{\delta F^\mu_{\ a}}{\delta e_\nu^{\ b}}.
\end{align}
Where we can derive the last term using \eqref{eqF}:
\begin{align}
\frac{\delta F^\mu_{\ a}}{\delta e_\nu^{\ b}}=\frac{1}{2}\frac{1}{((eV)^2)^2}\frac{\delta}{\delta e_\mu^{\ a}}(eV)^2\frac{\delta}{\delta e_\nu^{\ b}}(eV)^2-\frac{1}{2}\frac{1}{(eV)^2}\frac{\delta^2}{\delta e_\nu^{\ b}
\delta e_\mu^{\ a}}(eV)^2.
\end{align}
More explicitly,
\begin{align}
\notag &\frac{\delta F^\mu_{\ a}}{\delta e_\nu^{\ b}}=2F^\mu_{\ a} F^\nu_{\ b}+\eta_{ab}u^\mu u^\nu \!+\! \frac{1}{\sqrt{-(eV)^2}}\!\left[\frac{\delta V^\mu}{\delta e_\nu^{ \ b}}(eu)_a \!+\!\frac{\delta V^\nu}{\delta e_\mu^{\ a}}(eu)_b \!+\!\left(\!e\frac{\delta V}{\delta e_\nu^{ \ b}}\!\right)_{\! a} u^\mu \!+\!\left(\!e\frac{\delta V}{\delta e_\mu^{ \ a}}\!\right)_{\! b} u^\nu \! \right]\\
&+\frac{1}{\sqrt{-(eV)^2}}\left( e\frac{\delta^2 V}{\delta e_\mu^{\ a}\delta e_\nu^{\ b}}\right)\cdot (e u)+\frac{1}{-(eV)^2}\left(e\frac{\delta V}{\delta e_\mu^{ \ a}} \right)\cdot \left(e\frac{\delta V}{\delta e_\nu^{ \ b}} \right).
\end{align}

In flat space $F^\mu_{\ a}= u^\mu (\delta u)_a$, therefore the second variation simplifies to 
\begin{align}
\notag &2 \vev{\cT^\mu_{\ a} \cT^\nu_{\ b}}=\frac{1}{|e|}\frac{\delta^2 W}{\delta e_\mu^{\ a} \delta e_\nu^{\ b}}\\
\notag &=(\delta^\mu_{\ a} \delta^\nu_{\ b}-\delta^\mu_{\ b} \delta^\nu_{\ a} )p+ T\frac{\partial p}{\partial T}(\delta^\mu_{\ a} u^\nu (\delta u)_b+\delta^\nu_{\ b} u^\mu (\delta u)_a) + T\frac{\partial}{\partial T}\left(\! T\frac{\partial p}{\partial T}\!\right) u^\mu u^\nu (\delta u)_a (\delta u)_b \\
\notag &\!+\!T\frac{\partial p}{\partial T}\!\left[\! (\eta_{ab} \!+\! 2 (\delta u)_a (\delta u )_b ) u^\mu u^\nu \!+\!
\frac{1}{\sqrt{-V^2}}\!\left[\!\frac{\delta V^\mu}{\delta e^{\ b}_{\nu}}(\delta u)_a  
\!+\! \frac{\delta V^\nu}{\delta e^{\ a}_{\mu}}(\delta u)_b \! + \!
 u^\mu \left( \! \delta \frac{\delta V}{\delta e_\nu^{\ b}} \! \right)_{\!\! a} 
\!+\! u^\nu \left(\! \delta \frac{\delta V}{\delta e_\mu^{\ a}} \! \right)_{\!\! b} 
 \right]\!\right. \\
&\qquad \quad  \left. 
+\frac{1}{\sqrt{-V^2}}\left(\delta\frac{\delta^2 V}{\delta e_\mu^{\ a}\delta e_\nu^{\ b}} \right) \!\cdot\!(\delta u) +\frac{1}{-V^2}\left(\delta \frac{\delta V}{\delta e_\mu^{\ a}} \! \right) \!\cdot\! \left( \delta \frac{\delta V}{\delta e_\nu^{\ b}}\right) \right].
\end{align}
In order to check the Kubo formulas we will evaluate this expression in the static background, with the flat space vierbeins and $V^0=1$, $V^i=0$. The most interesting formulas are those involving a mixture of time and space indices, since this is where we expect to see the difference between Lifshitz and relativistic theories. The time-space correlators become
\begin{align}
\vev{T^0_{\ i} T^0_{\ j}} &=\frac{\varepsilon_0+p_0}{2}\left[\delta_{ij}+\frac{\delta V_i}{\delta e_0^{\ j}} +\frac{\delta V_j}{\delta e_0^{\ i}} +\delta_{kl}\frac{\delta V^k}{\delta e_0^{\ i}}\frac{\delta V^l}{\delta e_0^{\ j}}-\frac{\delta^2 V^0}{\delta e_0^{\ i}\delta e_0^{\ j}}\right],\\
\vev{T^0_{\ i} T^j_{\ 0}} &=-\frac{p_0}{2}\delta^i_j +\frac{\varepsilon_0+p_0}{2}\left[\frac{\delta V_i}{\delta e_j^{\ 0}} -\frac{\delta V^j}{\delta e_0^{\ i}} +\delta_{kl}\frac{\delta V^k}{\delta e_0^{\ i}}\frac{\delta V^l}{\delta e_j^{\ 0}}-\frac{\delta^2 V^0}{\delta e_0^{\ i}\delta e_j^{\ 0}}\right],\\
\vev{T^i_{\ 0} T^j_{\ 0}} &=\frac{\varepsilon_0+p_0}{2}\left[-\frac{\delta V^i}{\delta e_j^{\ 0}} -\frac{\delta V^j}{\delta e_i^{\ 0}} +\delta_{kl}\frac{\delta V^k}{\delta e_i^{\ 0}}\frac{\delta V^l}{\delta e_j^{\ 0}}-\frac{\delta^2 V^0}{\delta e_i^{\ 0}\delta e_j^{\ 0}}\right].
\end{align}
We have used the condition that in the static background
\begin{equation}
\frac{\delta V^0}{\delta e_\mu^{\ a}}=0.
\end{equation}
We will recover the right Kubo formulas for the relativistic case if
\begin{align}
V^i &= -\frac{1}{2}\frac{p_0}{\varepsilon_0+p_0} e_0^{\ i}-\frac{1}{2}\frac{\varepsilon_0}{\varepsilon_0+p_0}\delta^{ik}e_k^{\ 0},\\
V^0 &=1+\frac{p_0^2}{8(\varepsilon_0+p_0)^2}\delta_{ij}e_0^{\ i} e_0^{\ j}+\frac{\varepsilon_0^2}{8(\varepsilon_0+p_0)^2}\delta^{ij}e_i^{\ 0} e_j^{\ 0} 
+\left[\frac{\varepsilon_0-p_0}{2(\varepsilon_0+p_0)}+\frac{\varepsilon_0 p_0}{4(\varepsilon_0+p_0)^2}\right]\delta^i_j e_i^{\ 0} e_0^{\ j}.
\end{align}

For the Lifshitz case these expressions are modified. Let us use the notation
\begin{equation}\label{t00def}
\vev{T^0_{\ i}T^0_{\ j}}=t^{00}\delta_{ij}, \ \vev{T^i_{\ 0}T^0_{\ j}}=t^0_0\delta^i_j, \ \vev{T^i_{\ 0}T^j_{\ 0}}=t_{00}\delta^{ij}.
\end{equation}
Then,
\begin{align}\label{Vvierb}
V^i &= -\left[\frac{1}{2}-\frac{t^{00}}{\varepsilon_0+p_0}\right] e_0^{\ i}-\frac{t_{00}}{\varepsilon_0+p_0}\delta^{ik}e_k^{\ 0},\\
V^0 &=1+\frac{1}{2}\left[\frac{1}{2}-\frac{t^{00}}{\varepsilon_0+p_0}\right]^2\delta_{ij}e_0^{\ i} e_0^{\ j}+\frac{1}{2}\frac{t_{00}^2}{(\varepsilon_0+p_0)^2}\delta^{ij}e_i^{\ 0} e_j^{\ 0} +C^0_0\,\delta^i_j e_i^{\ 0} e_0^{\ j},
\end{align}
where
\begin{equation}
C^0_0=\frac{1}{2}-\frac{p_0+\frac{1}{2}t_{00}+ t^{00} +2 t^0_0}{\varepsilon_0+p_0}-\frac{t_{00} t^{00}}{(\varepsilon_0+p_0)^2}.
\end{equation}
We see that at least up to the level we have made our calculations, it is consistent to use the hydrodynamic generating functional formalism for Lifshitz theories. At the ideal level one can distinguish a Lifshitz theory from a relativistic theory with the same equation of state by putting the theories in curved spacetime. The equilibrium values of the velocities and conserved densities are different in both theories, as they are determined to leading order by the two-point functions (and to higher orders by higher correlation functions) and in particular they will exhibit a different temperature dependence. On the other hand, the form of the ideal hydrodynamic energy-momentum tensor in terms of energy, pressure and velocities is universal and no additional frame-independent terms are present.

%%%%%%%%%%%%%%%%%%%%%%%%%%%%%%%%%%%%%%%%%%%%%%%%%%%%%%%%%%%
\section{Ideal hydrodynamics from generating functional}
%%%%%%%%%%%%%%%%%%%%%%%%%%%%%%%%%%%%%%%%%%%%%%%%%%%%%%%%%%%

We will follow  the approach of \cite{Banerjee:2012iz,Jensen:2012jh} to derive the ideal hydrodynamic equations and the trace Ward identity from a generating functional. We couple the Lifshitz theory to a background vierbein and integrate out all degrees of freedom. The resulting generating functional must be invariant under the symmetries of the Lifshitz theory, thus by doing symmetry transformations of the sources one finds conditions on the generating functional that one can interpret as the hydrodynamic equations. In all this procedure the generating functional is taken to have a local dependence on the sources, which is appropriate for equilibrium configurations at finite temperature in the absence of long-range spatial correlations. The symmetry transformations are then generalized to local transformations. The non-relativistic form of general coordinate transformations in the rest frame were originally used in in \cite{2006AnPhy.321..197S} to constrain the effective action of unitary Fermi gases, and later generalized in \cite{Son:2008ye} and extended for $z\neq 2$ in \cite{Janiszewski:2012nb}.

The Lifshitz symmetry algebra consists of a non-relativistic scale transformation 
$D=-(z t\partial_t + x^i \partial_i)$, time translation $H=-\partial_t$, 
spatial translations $P_i=-\partial_i$ and rotations $M_{ij}= -x^i \partial_j + x^j \partial_i$. 
The corresponding commutation relations are 
\begin{align}
&[M_{ij}, M_{kl}] = \delta_{ik} M_{jl} + \delta_{jl} M_{ik} 
- \delta_{il} M_{jk} -\delta_{jk} M_{il}  \;, \\
&[M_{ij}, P_{k}] = \delta_{ik} P_{j} - \delta_{jk} P_{i}  \;, \\
&[D, P_i] = P_i  \;, \quad [D,H]=zH \;. 
\end{align}
The scale symmetry associated to the $D$ generator implies the trace Ward identity
\begin{equation}
z T^0_{\ 0}+\sum_i T^i_{\ i}=0.
\end{equation}
This algebra is appropriate in the rest frame of the fluid for a Lifshitz theory with $t^\mu=(1,\mathbf{0})$. However, when the fluid is moving (with constant velocity) we should use the Lifshitz theory in a different frame. As we have seen in the example of the free scalar, this amounts to changing the time-like vector $t^\mu\to u^\mu$. The symmetries of the Lifshitz theory are affected by the change of frame. The generators associated to translations and scale transformations become
\begin{equation}\label{lifshgen}
P^\parallel= u^\mu \partial_\mu, \ \ P^\perp_\mu=P_\mu^{\ \nu}\partial_\nu, \ \ D=z x^\mu u_\mu P^\parallel-x^\mu P^\perp_\mu .
\end{equation}
Where $P^{\ \nu}_{\mu}=\delta^{\ \nu}_{\mu}+u_\mu u^\nu$. Then, the momentum operators commute among themselves and
\begin{equation}
[D,P^\parallel]=z P^\parallel \ , \quad [D,P^\perp_\mu]=P^\perp_\mu \ .
\end{equation}
The Ward identity associated to $D$ becomes
\begin{equation}\label{wardu}
z T^\mu_{\ \nu} u_\mu u^\nu-T^\mu_{\ \nu} P_\mu^{\ \nu}=0 \ .
\end{equation}

For the background vierbein we will use the following parametrization:
\begin{equation}
e_0^{\ 0}=e^{-\alpha \Phi}, \ \ e_i^{\ 0}=-e^{-\alpha \Phi}B_i, \ \ e_0^{\ i}=0,
\end{equation}
and define the spatial metric
\begin{equation}
g_{ij}=\delta_{kl}e_i^{\ k}e_j^{\ l}.
\end{equation}
This is equivalent to the following form of background metric
\begin{align} \label{backgroundMetric}
ds^2 &=G_{\mu\nu}dx^\mu dx^\nu=-e^{-2\alpha \Phi}\left(dx^0-B_i dx^i\right)^2 
+ g_{ij}dx^idx^j  \;.
\end{align}
Under infinitesimal diffeomorphisms $x^\mu \to x^\mu+\xi^\mu$, the metric components transform as
\begin{equation}
\delta G_{\mu\nu}=\xi^\rho \partial_\rho G_{\mu\nu}+\partial_\mu\xi^\rho G_{\rho\nu}+\partial_\nu\xi^\rho G_{\mu\rho}.
\end{equation}
For general temporal ($\xi^0=f$) and spatial ($\xi^k$) diffeomorphisms the transformations are 
\begin{align}
\notag &\delta_\xi \Phi = \xi^\mu \partial_\mu \Phi 
+ \frac{1}{\alpha} \left(\dot \xi^k B_k - \dot f \right)\;,  \\
\notag &\delta_\xi B_i = \xi^\mu \partial_\mu B_i + (\dot \xi^k B_k -\dot f) B_i + \partial_i \xi^k B_k 
- \partial_i f + \dot \xi^k g_{ki} e^{2\alpha \Phi} \;, \\
&\delta_\xi g_{ij} = \xi^\mu \partial_\mu g_{ij} + \partial_i \xi^k g_{kj} + \partial_j \xi^k g_{ik} 
+ \dot \xi^k (B_i g_{kj} + B_j g_{ki})  \;.  \label{diffs}
\end{align} 
Where a dot denotes a time derivative $\dot f=\partial_t f$, etc. 

In the frame defined by $u^\mu$, the form of an infinitesimal scale transformation with parameter $\omega$ is
\begin{align}  \label{CoorScaleTr}
&\xi^\mu = \omega \left(- x^\mu +(z-1)u^\mu u_\alpha x^\alpha \right) \;.
\end{align}
The transformation of the background metric is then 
\begin{align}\label{scaletr}
&\delta_\omega \Phi = \xi^\alpha\partial_\alpha \Phi-\frac{\omega}{\alpha}\left[1-(z-1)(u^0-B_k u^k) u_0 \right]\;, \nonumber \\
&\delta B_i=\xi^\alpha\partial_\alpha B_i-(z-1)\omega\left[e^{2\alpha\Phi}g_{ik}u^k u_0-(u^0-B_k u^k)(u_0 B_i+u_i)\right]\;, \nonumber \\
&\delta g_{ij} =\xi^\alpha\partial_\alpha g_{ij}+\omega\left[2g_{ij}-(z-1) 2 u^k g_{(k}( u_{j)}+u_0 B_{j)}) \right]\;.
\end{align}

Which can be seen as a generalization of a Weyl transformation. Indeed, note that for the relativistic case $z=1$, we recover the usual Weyl transformation
\begin{align}
	\delta_\omega \Phi = -\frac{1}{\alpha} \omega \;,   \quad 
	\delta_\omega B_i = 0 \;, \quad 
	\delta_\omega g_{ij} = 2\omega g_{ij}  \;. 
\end{align}

We will slightly generalize the analysis by considering charged hydrodynamics with a single $U(1)$ charge. We will introduce as a source an background gauge field $A_\mu$.%
\footnote{One way to realize this is introducing an additional coordinate $\xi$ and add 
a term $ ( dy - A_0 dx^0 - A_i dx^i )^2$ to the metric \eqref{backgroundMetric}. 
The coordinate $y$ does not carry a non-trivial scaling dimension. 
}
The change of the gauge field under gauge transformations ($\lambda$) and diffeomorphisms is
\begin{equation}
\delta_\xi A_\mu=\xi^\alpha \partial_\alpha A_\mu+\partial_\mu\xi^\alpha A_\alpha-\partial_\mu\lambda.
\end{equation}
Or, distinguishing between temporal and spatial diffeomorphisms
\begin{align}
&\delta_\xi A_0 = \xi^\mu \partial_\mu A_0 + \dot f A_0 + \dot \xi^k A_k - \dot \lambda \;,  \nonumber \\ 
&\delta_\xi A_i = \xi^\mu \partial_\mu A_i + \partial_i f A_0 + \partial_i \xi^k A_k - \partial_i \lambda \;, 
\end{align}
in addition to \eqref{diffs}. 
In the frame with non-trivial $u^\mu$, the infinitesimal scaling transformation of the background gauge fields are 
\begin{align} 
&\delta_\omega A_\mu= \xi^\alpha\partial_\alpha A_\mu + \omega\left[A_\mu-(z-1)u_\mu u^\alpha A_\alpha\right].
\end{align}
in addition to \eqref{scaletr}.

The generating functional depending on the background metric and gauge fields,  frame, temperature and chemical potential is 
\begin{align} \label{generatingW}
W=\int d^{d+1}x\sqrt{g}e^{-\alpha \Phi} p(T,\mu) \;,
\end{align}
where
\begin{align}
&T=\frac{T_0}{\sqrt{-G_{\mu\nu} V^\mu V^\nu}} = 
\frac{T_0 e^{\alpha  \Phi}}{\sqrt{(V^0-B_iV^i)^2 - e^{2\alpha \Phi} g_{ij} V^iV^j}} \;,\\ 
&\mu=\frac{A_\mu V^\mu  }{\sqrt{-G_{\mu\nu} V^\mu V^\nu}} = 
\frac{e^{\alpha \Phi}A_\mu V^\mu}{\sqrt{(V^0-B_iV^i)^2 - e^{2\alpha \Phi} g_{ij} V^iV^j}} \;, 
\end{align}
The vector $V^\mu$ is proportional to the frame velocity $u^\mu$ but it is not of unit norm, the relation between the two is
\begin{equation}
u^\mu=\frac{V^\mu}{\sqrt{(V^0-B_iV^i)^2 - e^{2\alpha \Phi} g_{ij} V^i V^j }}.
\end{equation}
We will now derive the hydrodynamic equations by imposing the symmetry of the generating functional under translations and scale transformations. Under a general transformation of the background fields
\begin{equation}
\delta_\xi W=\delta_\xi g_{ij}\frac{\delta W}{\delta g_{ij}}+\delta_\xi B_i\frac{\delta W}{\delta B_i}+\delta_\xi \Phi \frac{\delta W}{\delta \Phi}+\delta_\xi A_\alpha \frac{\delta W}{\delta A_\alpha}.
\end{equation}
When the sources are set to zero ($\Phi=0$, $B_i=0$, $g_{ij}=\delta_{ij}$, $A_0 u^0=\mu$), the variation of the generating functional is simply
\begin{align}
& \delta_\Phi W = \alpha \left[ (\varepsilon+p)u^0 u^0+\eta^{00}p\right]\;,\\
&\delta_{B_i} W = (\varepsilon+p) u^i u^0\;,\\
&\delta_{g_{ij}}W = \frac{1}{2}\delta^{ij} p+\frac{1}{2}(\varepsilon+p)u^i u^j\;,\\
&\delta_{A_\alpha} W = \rho u^\alpha \;.
\end{align}
In order to get these expressions we have used the relations
\begin{equation}
s=\frac{\partial p}{\partial T}, \ \ \rho=\frac{\partial p}{\partial \mu}, \ \ Ts+\mu\rho=\varepsilon+p.
\end{equation}

Invariance under temporal and spatial diffeomorphisms \eqref{diffs} imply the equations
\begin{align}
& \partial_\mu T^{\mu 0}=-\rho u^\nu \partial_\nu A^0, \\
& \partial_\mu T^{\mu i}=-\rho u^\nu \partial_\nu A^i = 0.
\end{align}
where the energy-momentum tensor is
\begin{equation}
T^{\mu\nu}=(\varepsilon+p)u^\mu u^\nu+ p \eta^{\mu\nu}.
\end{equation}
The conservation equations thus take the same form as in a relativistic theory.

Invariance under the Weyl transformations \eqref{scaletr} imposes the following condition 
%(for a constant chemical potential)
\begin{align}
-z (\varepsilon-\mu\rho) + dp = 0 \;.
\end{align}
For $\mu=0$ this is the usual equation of state for the theory with Lifshitz scale invariance. Note that the way we have defined it, it is independent of the frame and in fact it agrees with the Ward identity \eqref{wardu}. When $\mu\neq 0$ the Ward identity is modified because the chemical potential breaks scale invariance. Note however that the temperature dependence of the pressure is still the same, since
\begin{equation}
0=-z (\varepsilon-\mu\rho) + dp=-z(Ts -p)+d p=-zT\frac{\partial p}{\partial T}+(d+z)p.
\end{equation}
Integrating this equation we find that $p\propto T^{\frac{z+d}{z}}$ even when the chemical potential is non-zero.

We can generalize this analysis to include higher order terms in a derivative expansion of the sources, which appear beyond the ideal order in the hydrodynamic expansion. Since the method is valid for equilibrium configurations, only non-dissipative terms are captured this way. In the original works  \cite{Banerjee:2012iz,Jensen:2012jh} the approach using the generating functional was useful to derive relations between transport coefficients at the same or different orders. It is an interesting problem that we leave for future work to find how the relations are modified in the non-relativistic theory, and whether they still match with the derivation using an entropy current.

%%%%%%%%%%%%%%%%%%%%%%%%%%%%%%%%%%%%%%%%%%%%%%%%%%%%%%%%%%%
\section{Ideal hydrodynamics from fluid/gravity correspondence}
%%%%%%%%%%%%%%%%%%%%%%%%%%%%%%%%%%%%%%%%%%%%%%%%%%%%%

A very successful application of gauge/gravity dualities has been the map between hydrodynamic equations of motion and Einstein equations in black hole geometries \cite{Bhattacharyya:2008jc}. Among the theories where the fluid/gravity correspondence has been derived are non-relativistic conformal theories \cite{Rangamani:2008gi}, which have a dynamical exponent $z=2$. Here we will extend the correspondence to other dynamical exponents. Our goal is to show that the form of the ideal energy-momentum tensor and the equation of state take the same form as we have derived in the previous sections. In principle we could extend the analysis to include higher derivative terms (dissipative and non-dissipative) by doing a systematic expansion of the metric and bulk fields, this is an interesting problem that we leave as future work. 

We will derive the hydrodynamic equations of the dual field theory following the method of projecting  Einstein's equations on the horizon \cite{Eling:2009pb,Eling:2009sj,Eling:2010hu} for a particular model with Lifshitz solutions proposed in \cite{Charmousis:2010zz,Gouteraux:2011ce}. We find the usual Navier-Stokes equations
\begin{equation}
\left(\varepsilon+p\right)u^\alpha\partial_\alpha u_\nu+ P_\nu^\alpha\partial_\alpha p =0.
\end{equation}
Where the energy and the pressure satisfy the Lifshitz equation of state $-z\varepsilon+d p=0$ and have the expected scaling with the temperature $\varepsilon \sim T^{\frac{d+z}{z}}$. 
The Lifshitz energy-momentum tensor at rest frame has been computed in \cite{Ross:2009ar,Ross:2011gu,Korovin:2013nha}. 

\subsection{Gravitational background}

Metrics with Lifshitz scaling were first constructed and proposed as holographic duals of Lifshitz theories in \cite{Kachru:2008yh}. Black hole geometries were first found analytically in \cite{Taylor:2008tg} and numerically in \cite{Danielsson:2009gi} and also analytically in \cite{Mann:2009yx,Pang:2009ad,Bertoldi:2009vn,Balasubramanian:2009rx}. 
As a first step we generalize such solutions by constructing a black-brane solution at a constant velocity using coordinate transformations. 

At zero temperature, the original Lifshitz metric is
\begin{equation}\label{lifshmet}
ds^2=-r^{2z}dt^2+\frac{dr^2}{r^2}+r^2 \delta_{ij} dy^i dy^j,
\end{equation}
which has the isometry
\begin{equation}\label{isomlifsh}
r\to \lambda r, \ \ t\to \lambda^{-z}t, \ \ y^i\to \lambda^{-1}y^i.
\end{equation}
Let us consider a black-hole metric of the form
\begin{equation}\label{bhmetric}
ds^2=-F(r)dt^2+H(r)^2\frac{dr^2}{F(r)}+G(r) \delta_{ij} dy^i dy^j
\end{equation}
where $F(r_H)=0$. In a Lifshitz black hole solution (in a 3+1 dimensional bulk) the functions appearing in the metric are $H(r)=r^{-z-1}$, $F(r)=r^{2z} f(r)$, $G(r)=r^{2}$ and $f(r)=1-\left(\frac{r_H}{r} \right)^{2+z}$.

We can do a change of coordinates
\begin{equation}
t=\tilde{t}+r_*(r),
\end{equation}
where
\begin{equation}
\frac{d r_*}{dr}=\frac{H(r)}{F(r)}
\end{equation}
Then, the metric becomes
\begin{equation}
ds^2=-2H(r) d\tilde{t} dr-F(r)d\tilde{t}^2+G(r) \delta_{ij} dy^i dy^j
\end{equation}
If the geometry is sourced by a matter energy-momentum tensor (with $T_{tr}=0$), the components in the $t$, $r$ directions are changed as
\begin{equation}
\tilde{T}_{\tilde{t}\tilde{t}}= T_{tt},
\end{equation}
\begin{equation}
\tilde{T}_{\tilde{t}r}=\frac{H(r)}{F(r)} T_{tt},
\end{equation}
\begin{equation}
\tilde{T}_{rr}=T_{rr}+\left(\frac{H(r)}{F(r)}\right)^2 T_{tt}.
\end{equation}

We can now do another coordinate transformation
\begin{equation}
\tilde{t}=u_\mu x^\mu, \ \ y^i=x^i+\gamma \beta^i x^0+(\gamma-1)\frac{\beta^i\beta^j}{\beta^2}x_j.
\end{equation}
Where $u_\mu=\gamma (1,\beta^i)$, and $\gamma^2=1/(1-\beta^2)$, $\beta^2=\beta_i\beta^i$. One can check that
\begin{equation}
\delta_{ij}\frac{\partial y^i}{\partial x^\mu}\frac{\partial y^j}{\partial x^\nu}=\eta_{\mu\nu}+u_\mu u_\nu=P_{\mu\nu}
\end{equation}
Then, the metric becomes
\begin{equation}\label{metricVelocity}
ds^2=-2H(r) u_\mu dx^\mu dr-F(r)u_\mu u_\nu dx^\mu dx^\nu+G(r) P_{\mu\nu}dx^\mu dx^\nu.
\end{equation}
This expression is our black brane solution for constant velocities. Note that the functions that depend on the radial coordinate are arbitrary, so in particular we can choose them to be those of Lifshitz black holes. When the velocities are not constant it is necessary to correct the metric by terms depending on the derivatives of the velocity, in order to ensure that the Einstein's equations are satisfied and the solution is regular. 

If we take \eqref{lifshmet} as starting point, the boosted brane solution is
\begin{equation}
ds^2=-2r^{z-1} u_\mu dx^\mu dr-r^{2z}u_\mu u_\nu dx^\mu dx^\nu+r^2 P_{\mu\nu}dx^\mu dx^\nu.
\end{equation}
Note that this is not invariant under the transformation \eqref{isomlifsh}. However, since a change of coordinates cannot make the isometry disappear, it must take a new form. Indeed, one can check that the metric is invariant under
\begin{equation}
r\to \lambda r, \ \ x^\mu\to -\lambda \left(P^\mu_{\ \alpha}x^\alpha-z u^\mu u_\alpha x^\alpha\right).
\end{equation}
Note that for $z=1$ the transformation is still the same, but not for general $z$. In terms of the components of a Killing vector
\begin{equation}
\xi^r= r, \ \ \xi^\mu=-P^\mu_{\ \alpha}x^\alpha+z u^\mu u_\alpha x^\alpha.
\end{equation}
In the dual field theory $\xi^\mu$ should map to a symmetry, this is precisely the scaling symmetry generated by the velocity-dependent $D$ in \eqref{lifshgen}. 
 
Regarding the matter fields, in the boosted brane solution the components of the matter energy-momentum tensor are
\begin{equation}
\hat{T}_{rr}=\tilde{T}_{rr},
\end{equation}
\begin{equation}
\hat{T}_{\mu r}=u_\mu \tilde{T}_{\tilde{t}r},
\end{equation}
\begin{equation}
\hat{T}_{\mu\nu}=u_\mu u_\nu \tilde{T}_{\tilde{t}\tilde{t}}+P_{\mu\nu} T_{11}
\end{equation}
where we have used $T_{ij}=T_{11}\delta_{ij}$. The changes of variables are the same for the components of the Ricci tensor.

If there are background scalar fields depending on the radial coordinate, their profile $\phi(r)$ is not affected by the change of coordinates. For gauge fields we will have
\begin{equation}
\cA=A_t dt+A_r dr =A_t\left(d\tilde{t}+\frac{H}{F}dr\right)+A_r dr=A_t u_\mu dx^\mu+\left(A_r+\frac{H}{F}\right)dr
\end{equation}
The conventions for the black brane are that $A\propto u_\mu dx^\mu$ is non-zero at the horizon. We should then make $A_t$ non-zero at the horizon and
\begin{equation}
A_r=-\frac{H}{F}.
\end{equation} 
Since $H$ and $F$ are only functions of $r$, this choice does not affect to the field strengths or the equations of motion. In the static case the $A_t\neq 0$  condition is not regular and $A_r\propto 1/F$ is singular, so there is no smooth mapping. Presumably is because the future and past horizons sit at the same value of $r$ in the static case and in the black brane they are separated. The field strengths are (allowing a space-time dependence on $u_\mu$ and $A_t$)
\begin{equation}
\cF=d\cA=A_t' u_\mu dr\wedge dx^\mu+ 2 \partial_{[\mu} A_t u_{\nu]}dx^\mu \wedge dx^\nu.
\end{equation}

\subsection{Lifshitz model}

The action for the effective model proposed in \cite{Charmousis:2010zz,Gouteraux:2011ce} consists of Einstein gravity coupled to Maxwell's gauge fields $F_{\mu\nu} $ and a scalar `dilaton' $\phi$
\begin{equation}\label{actionlifsh}	
	  	S =\int  d ^{d+2}x~\sqrt{-g}\left[R- \frac{Z(\phi)}{4}F_{AB}F^{AB}
  		-\frac{1}{2}(\partial\phi)^2 + V(\phi) \right],  
\end{equation}
where $ A, B = (r, \mu)$, $d$ is a number of spatial dimensions, and we focus on $d=2$. 
Einstein's equations are
\begin{equation}
R_{AB}-\frac{1}{2}g_{AB} R=T_{AB}^s+T_{AB}^M,
\end{equation}
where the energy-momentum tensor for the scalar field is
\begin{equation}
T_{AB}^s=\frac{1}{2}\partial_A\phi\partial_B\phi-\frac{1}{4}g_{AB}\left[(\partial\phi)^2 -2 V(\phi)\right],
\end{equation}
and the Maxwell's part is
\begin{equation}
T_{AB}^M=\frac{Z(\phi)}{2}\left(g^{CD}F_{AC}F_{BD}-\frac{1}{4}g_{AB}F^2\right).
\end{equation}
For the matter fields we are interested in Maxwell's equations 
\begin{equation}
\partial_A\left(\sqrt{-g}Z(\phi)g^{AB}g^{CD}F_{BD}\right)=0.
\end{equation}

These set of equations admit black hole solutions \eqref{bhmetric} with two independent parameters, 
Lifshitz scaling exponent $z$ and hyperscaling violation exponent $\theta$. Explicitly 
\cite{Charmousis:2010zz,Kim:2012pd},  
\begin{align}\label{metricfuncs}
	&H(r)=r^{-z-1-2\theta/d}, \quad F(r)=r^{2z-2\theta/d} f(r), \quad G(r)=r^{2-2\theta/d}. 
\end{align}
Where $f(r)=1-\left(\frac{r_H}{r} \right)^{d+z-\theta}$. The solutions for the scalar and background gauge fields are
\begin{equation}
e^{\phi(r)}= r^s, \ \ A_t=\sqrt{\frac{2(z-1)}{z+2-\theta}} r^{2+z-\theta} f(r), 
\end{equation}
where $ s=\pm \sqrt{4z-4 + \theta^2-2z\theta} $, 
while the coupling $Z(\phi)$ and a scalar potential $V(\phi)$ are 
\begin{equation}
Z(\phi)= e^{\frac{4-\theta}{s}\phi}, \ \ V(\phi)= (2+z-\theta)(1+z-\theta) e^{-\frac{\theta}{s}\phi}. 
\end{equation}
This solution can be viewed as a direct generalization of AdS black hole with dynamical $z$ and 
hyperscaling violation $\theta$ exponents. 

We can do a coordinate transformation to the metric \eqref{metricVelocity} and after that introduce a spacetime dependence in the velocities, $r_H$, the gauge field and the scalar.

\subsection{Projection of equations of motion on the horizon}

We will project with the normal to the horizon $\ell^A$ 
\begin{equation}
\ell^A=(0,u^\mu),
\end{equation}
and evaluate the projected equations at the horizon. For this, we will need the inverse metric
\begin{equation}
g^{AB}=\left(
\begin{array}{cc}
\frac{F}{H^2} & \frac{u^\nu}{H} \\
\frac{u^\mu}{H} & \frac{1}{G}P^{\mu\nu}
\end{array}
\right).
\end{equation}
The details of the projection are collected in Appendix \ref{sec:curvature}. Note that at the horizon $g^{rr}=F/H^2=0$ so $\ell^A$ is indeed a normal vector.

The simplest equation is the current conservation obtained from Maxwell's equations
\begin{equation}
\partial_\mu(\rho u^\mu)=0,
\end{equation}
where
\begin{equation}
\rho=-\frac{ Z(\phi) G^{d/2}}{16\pi H}A_t',
\end{equation}
evaluated at the horizon. The factor of $16\pi $ is arbitrary and is fixed for convenience.

The scalar energy-momentum tensor does not contribute to the ideal conservation equations, but it will contribute to bulk viscosity terms in the hydrodynamic equations as shown in \cite{Eling:2011ms}. The contribution from the Maxwell's fields to hydrodynamic equations are
\begin{equation}
T_{\mu B}^M\ell^B=\frac{Z(\phi)}{2H}\left(2 A_t'u^\nu\partial_{[\mu}A_t u_{\nu]}\right)=\frac{2\pi \rho}{s} \left(P_\mu^\alpha\partial_\alpha A_t+A_t u^\alpha\partial_\alpha u_\mu\right)
\end{equation}
where we have used that $G^{d/2}=4s$ gives the entropy density. Projecting with $u^\mu$ this term vanishes, so there is no contribution to the entropy current. Projecting with $P^\mu_\nu$ we get the same equation.

So far our expressions are valid for a general action of the form \eqref{actionlifsh}, but we will need to be more concrete now and we will use \eqref{metricfuncs}. The surface gravity determines the temperature
\begin{equation}
\kappa=2\pi T=\frac{1}{2}(z+d-\theta)r_H^z \equiv 2\pi b r_H^z.
\end{equation}
For convenience we will define $T_b=T/b$, which fixes
\begin{align}
G=T_b^{\frac{2-2\theta/d}{z}},\ \
 s=\frac{1}{4}r_H^{(d-\theta)/z}=\frac{1}{4}T_b^{(d-\theta)/z},\ \
H=T_b^{-\frac{1+z+2\theta/d}{z}}.
\end{align}
Then, $\partial_\mu G/G=\frac{2-2\theta/d}{z}\partial_\mu T_b/T_b$. 

The projection of Einstein's equations evaluated on these solutions gives
\begin{equation}\label{einsteq}
R_{\mu \nu} \ell^\nu=-2\pi b\left[\frac{(d-\theta)}{z}u_\mu u^\alpha\partial_\alpha T_b+T_b\partial_\alpha P^\alpha_\mu+P_\mu^\alpha\partial_\alpha T_b  \right].
\end{equation}
Projecting with $u^\mu$ we get
\begin{equation}
2\pi b\left[\frac{(d-\theta)}{z} u^\alpha\partial_\alpha T_b+T_b\partial_\alpha u^\alpha \right]=
\frac{2\pi b}{T_b^{\frac{d-\theta}{z}-1}} \partial_\alpha(T_b^{\frac{d-\theta}{z}} u^\alpha) 
=\frac{8\pi b}{T_b^{\frac{d-\theta}{z}-1}}  \partial_\alpha(s u^\alpha).
\end{equation}
Since the matter fields do not contribute to leading order, the entropy current is conserved
\begin{equation}
\partial_\mu (s u^\mu)=0.
\end{equation}

Projecting now \eqref{einsteq} with $P^\mu_\nu$, we get
\begin{equation}
-2\pi b\left[T_b u^\alpha\partial_\alpha u_\nu+P_\nu^\alpha\partial_\alpha T_b \right].
\end{equation}
Together with the energy-momentum tensor of matter we have 
\begin{equation}
\left(1+ \frac{\rho}{ Ts}A_t\right)u^\alpha\partial_\alpha u_\nu+P_\nu^\alpha\partial_\alpha \ln T +\frac{\rho}{Ts} P_\nu^\alpha\partial_\alpha A_t=0.
\end{equation}
If we define the chemical potential as 
\begin{equation}
A_t(r_H)=\mu,
\end{equation}
then, multiplying by a factor of $Ts$
\begin{equation}
\left(Ts+ \mu \rho\right)u^\alpha\partial_\alpha u_\nu+Ts P_\nu^\alpha\partial_\alpha \ln T +\rho P_\nu^\alpha\partial_\alpha \mu=0.
\end{equation}
These are the ordinary hydrodynamic equations for a charged fluid. It becomes more clear if we use the thermodynamic identities
\begin{equation}
dp=\rho d\mu+sdT, \ \ \varepsilon+p=Ts+\mu\rho.
\end{equation}
The equations become the relativistic Navier-Stokes equations
\begin{equation}
\left(\varepsilon+p\right)u^\alpha\partial_\alpha u_\nu+ P_\nu^\alpha\partial_\alpha p =0.
\end{equation}
The exponents $z$ and $\theta$ of the theory are manifested in the dependence of the pressure, energy, charge and entropy density on the temperature and chemical potential.

%%%%%%%%%%%%%%%%%%%%%%%%%%%%%%%%%%%%%%%%%%%%%%%%%%%%%%%%%%%
\section{First order asymmetric dissipative terms}
%%%%%%%%%%%%%%%%%%%%%%%%%%%%%%%%%%%%%%%%%%%%%%%%%%%%%%%%%%

The breaking of Lorentz invariance implies that the energy-momentum tensor is not necessarily symmetric. We have seen from the calculation in free field theories that no asymmetric terms are expected in the hydrodynamic energy-momentum tensor at the ideal level. However, such terms could appear at higher orders in derivatives, although they can be constrained by physical requirements such as the second law of thermodynamics in its local form. In a previous work \cite{Hoyos:2013eza} we found the asymmetric terms possible to first viscous order, we will give here a more detailed presentation including a conserved current and the Kubo formulas for the new transport coefficients. 

The energy-momentum tensor in the Landau frame takes the form
\begin{equation}
T^{\mu\nu}=\varepsilon u^\mu u^\nu +p P^{\mu\nu}+\pi_S^{(\mu\nu)}+\pi_A^{[\mu\nu]} +(u^\mu\pi_A^{[\nu\sigma]}+u^\nu\pi_A^{[\mu\sigma]})u_\sigma.
\end{equation}
Where we impose on the symmetric part $\pi^{(\mu\nu)}_S u_\nu=0$ and the last term ensures that the condition
\begin{equation}
T^{\mu\nu} u_\nu=-\varepsilon u^\mu,
\end{equation}
is satisfied. To first order in derivatives the only possible contributions to $\pi_S$ are the shear and bulk viscosities
\begin{equation}
\pi_S^{(\mu\nu)}=-\eta^{\mu\nu\alpha\beta}\partial_\alpha u_\beta=-\eta P^{\mu\alpha} P^{\nu\beta} \Delta_{\alpha\beta}-\frac{\zeta}{d} P^{\mu\nu}\partial_\alpha u^\alpha,
\end{equation}
where $\eta$ and $\zeta$ are the shear and bulk viscosities respectively and the shear tensor is defined as
\begin{equation}
\Delta_{\alpha\beta}=2\partial_{(\alpha} u_{\beta)}-\frac{2}{d}P_{\alpha\beta}(\partial_\sigma u^\sigma).
\end{equation}

The constitutive relation of the conserved current is
\begin{equation}
J^\mu=\rho u^\mu +\nu^\mu,
\end{equation}
where we impose the condition $\nu^\mu u_\mu=0$.

The divergence of the entropy current is
\begin{align}
\notag 0 &=\partial_\mu T^{\mu\nu}u_\nu +\mu \partial_\mu J^\mu \\
\notag &=-T\partial_\mu j_s^\mu \!+\! \partial_\mu(\pi_A^{[\mu\nu]})u_\nu
\!+\! \partial_\mu(u^\mu\pi_A^{[\nu\sigma]}u_\sigma)u_\nu \!-\! \partial_\mu(\pi_A^{[\mu\sigma]})u_\sigma \!-\!  \pi_A^{[\mu\sigma]} \partial_\mu u_\sigma \!+\! \mu\partial_\mu \nu_A^\mu \!+ \cdots\\
&=-T\partial_\mu j_s^\mu-\pi_A^{[\mu\sigma]}(\partial_{[\mu} u_{\sigma]}-u_{[\mu} u^\alpha \partial_\alpha u_{\sigma]})+\mu\partial_\mu \nu^\mu+\cdots.
\end{align}
The dots denote positive-definite contributions from the shear and bulk viscosities that do not affect to the analysis. The entropy current is defined as
\begin{equation}
j_s^\mu=s u^\mu-\frac{\mu}{T}\nu^\mu.
\end{equation}

If the chemical potential is zero, in order to have a positive quantity, 
\begin{equation}
\pi_{A}^{[\mu\nu]}=-\alpha^{\mu\nu\alpha\beta}(\partial_{[\alpha} u_{\beta]}-u_{[\alpha} u^\rho \partial_\rho u_{\beta]})
\end{equation}
where $\alpha^{\mu\nu\sigma\rho}$ contains all possible transport coefficients to first dissipative order. It must also satisfy the condition, for an arbitrary real tensor $\tau_{\mu\nu}$,
\begin{equation}
\tau_{\mu\nu}\alpha^{\mu\nu\sigma\rho}\tau_{\sigma\rho} \geq 0 \ .
\end{equation}
The condition that boost but not rotational invariance is broken with respect to the rest frame of the fluid imposes the condition
\begin{equation}
P_{\alpha \mu} \pi_A^{[\mu\nu]} P_{\nu \beta}=0.
\end{equation}
This implies that the antisymmetric term should take the form
\begin{equation}
\pi^{[\mu\nu]}_A=u^{[\mu}V_A^{\nu]},
\end{equation}
where one can take $V_A^\nu u_\nu=0$ without loss of generality. This restricts the form of the transport coefficients $\alpha^{\mu\nu\alpha\beta}\sim u^{[\mu}P^{\nu][\beta}u^{\alpha]}$, and for the normal fluid it makes $\alpha^{\mu\nu\alpha\beta}\partial_{[\alpha}u_{\beta]}\sim u^{[\mu} a^{\nu]}$, where the acceleration is defined as $a^\mu=u^\alpha\partial_\alpha u^\mu$. This leads to a single transport coefficient
\begin{equation}
\pi^{[\mu\nu]}_A = -\alpha u^{[\mu}a^{\nu]}, \ \ \alpha\geq 0.
\end{equation}
When the chemical potential is non-zero there are two additional possible transport coefficients. One of them does not satisfy the Onsager relation and is {\em dissipationless}, while the other is dissipative. The equation for the entropy current is
\begin{equation}
\partial_\mu j_s^\mu=-\frac{1}{T}a^\alpha V_{A\,\alpha}-\nu^\mu\partial_\mu\left(\frac{\mu}{T}\right) +\cdots.
\end{equation}
Where the entropy current is defined as
\begin{equation}
j_s^\mu=s u^\mu-\frac{\mu}{T}\nu^\mu.
\end{equation}
We now expand the dissipative terms as
\begin{align}
V_A^\mu= -T\alpha_1 a^\mu-T\alpha_2 P^{\mu\nu}\partial_\nu\left(\frac{\mu}{T}\right),\\
\label{nua} \nu^\mu= -\alpha_3 a^\mu-\alpha_4 P^{\mu\nu}\partial_\nu\left(\frac{\mu}{T}\right).
\end{align}
Then, the equation for the entropy current becomes
\begin{equation}
\partial_\mu j_s^\mu =\left(
\begin{array}{cc} a^\mu  & P^{\mu\nu}\partial_\nu\left(\frac{\mu}{T}\right)
\end{array} 
\right) \left(\begin{array}{cc} \alpha_1 & \alpha_2 \\ \alpha_3 & \alpha_4 \end{array}\right) \left(\begin{array}{c} a_\mu  \\ P_\mu^{\ \lambda}\partial_\lambda\left(\frac{\mu}{T}\right) \end{array} \right).
\end{equation}
If we write $\alpha_2=C+\alpha'$, $\alpha_3=-C+\alpha'$, the dependence on $C$ drops from the equation, so it corresponds to a dissipationless transport coefficient, but it would be forbidden if we impose the Onsager relation. The other three  dissipative coefficients are $\alpha_1=\alpha/T$ (that is the same as in the neutral case), $\alpha'$ and the coefficient $\alpha_4=\sigma T$ that can be identified with the ordinary conductivity. The positivity conditions on the coefficients are 
\begin{equation}
\alpha\sigma\geq (\alpha')^2, \ \ \alpha\geq 0, \ \ \sigma\geq 0
\end{equation}

%%%%%%%%%%%%%%%%%%%%%%%%%%%%%%%%%%%%%%%%%%%%%%%%%%%%%%%%%%%
\subsection{Kubo formulas}
%%%%%%%%%%%%%%%%%%%%%%%%%%%%%%%%%%%%%%%%%%%%%%%%%%%%%

We will derive Kubo formulas for the new transport coefficients assuming that the Onsager relation is satisfied. To first order in the derivative expansion we have found the following asymmetric contributions to the energy-momentum tensor
\begin{equation}
T^{\mu\nu}u_\mu P_{\nu\alpha}=2\pi_A^{[\mu\nu]}u_\mu P_{\nu\alpha}, \ \ T^{\mu\nu} P_{\mu\alpha}u_\nu=0,
\end{equation}
that also enter in the current through $\nu$ in \eqref{nua}.
Expanding around the equilibrium configuration $u^\mu\simeq (1,\beta^i)$ to linear order in the velocities,
\begin{align}
&T_{0i}\simeq \alpha \partial_0 \beta_i+T\kappa \partial_i\left(\frac{\mu}{T} \right),\\
&j^i = \rho \beta^i-\sigma T \partial^i \left(\frac{\mu}{T} \right)-\alpha' \partial_0 \beta^i.
\end{align}

By expanding around flat spacetime, the dependence of velocity $\beta_i\simeq \delta_{ij} V^j$ on the background vierbein is as determined in \eqref{Vvierb} to leading order in derivatives. Recall that the two-point functions of the energy-momentum tensor are
\begin{equation}
\vev{T_{0i} T^j_{\ 0}}=\frac{\delta \vev{T_{0i}}}{\delta e_j^{ \ 0}}, \ \ \vev{T_{0i} T^0_{\ j}}=\frac{\delta \vev{T_{0i}}}{\delta e_0^{ \ j}}. 
\end{equation}
The mixed correlators with the current are
\begin{equation}
\vev{j^i T^j_{\ 0}}=\frac{\delta \vev{j^i}}{\delta e_j^{ \ 0}}, \ \ \vev{j^i T^0_{\ j}}=\frac{\delta \vev{j^i}}{\delta e_0^{ \ j}}. 
\end{equation}
Then, it is straightforward to derive the Kubo relations for the new transport coefficients after doing a Fourier transformation of the correlators\footnote{Note that already at the ideal level there is a correction depending on time derivatives of the vierbein to the static velocity. From the equation $\partial_t \delta\rho+\rho\partial_i \beta^i=0$ and the expressions \eqref{Vvierb} we see that it must be at least quadratic in the mixed components of the vierbeins, so it vanishes from the two-point function in flat spacetime.}
\begin{align}
\alpha &=  \lim_{\omega \to 0}\; \frac{1}{i\omega B} \vev{T_{0i} T^i_{\ 0}}(\omega,\mathbf{k}=0), \\
\alpha &= \lim_{\omega \to 0}\; \frac{1}{i\omega A} \vev{T_{0i} T^0_{\ i}}(\omega,\mathbf{k}=0), \\
\alpha' &=  -\lim_{\omega \to 0}\; \frac{1}{i B}\frac{\partial}{\partial \omega} \vev{j^i T^i_{\ 0}}(\omega,\mathbf{k}=0), \\
\alpha' &= -\lim_{\omega \to 0}\; \frac{1}{i A} \frac{\partial}{\partial \omega}\vev{j^i T^0_{\ i}}(\omega,\mathbf{k}=0). 
\end{align}
Where we have defined the coefficients
\begin{equation}
A=\frac{1}{2}-\frac{t^{00}}{\varepsilon_0+p_0}, \ \ B=\frac{t_{00}}{\varepsilon_0+p_0},
\end{equation}
and $t^{00}$, $t_{00}$ are the zero frequency two-point functions \eqref{t00def}.

%%%%%%%%%%%%%%%%%%%%%%%%%%%%%%%%%%%%%%%%%%%%%%%%%%%%%%%%%%%
\section{Discussion}
%%%%%%%%%%%%%%%%%%%%%%%%%%%%%%%%%%%%%%%%%%%%%%%%%%%%%

In the first couple of sections we have computed the energy momentum tensor at non-zero temperature of a free scalar. An obvious extension will be to add interactions and compute hydrodynamic transport coefficients using Kubo formulas. Of particular interest are the transport coefficients associated to the breaking of Lorentz invariance that we have found in the last section.

We have shown how to obtain the ideal energy-momentum tensor in a charged fluid using the generating functional and a holographic dual. In relativistic theories in some cases a current is conserved only up to a quantum anomaly, but nevertheless it can be included in the hydrodynamic description and the anomaly produces interesting effects in the motion of the fluid \cite{Erdmenger:2008rm,Banerjee:2008th,Son:2009tf,Neiman:2010zi,Landsteiner:2011cp}. Even if the fluid is not charged a conformal anomaly determines the equation of state \cite{Jensen:2012kj,Eling:2013bj,Banerjee:2013fqa}. Fluids with Lifshitz scaling may exhibit analogous properties, as there can be both axial anomalies 
\cite{Bakas:2011nq,Bakas:2011uf} and Weyl anomalies \cite{Adam:2009gq,Gomes:2011di, Baggio:2011ha,Griffin:2011xs}. We should remark that this is not just a formal observation about some models with Lifshitz scaling, anomalies may appear in the effective description of ordinary Fermi liquids \cite{Son:2012wh}. 

Another interesting direction that can also be pursued is to extend the hydrodynamic description to superfluids (see e.g.\cite{Landau}). This is motivated by the apparent existence of a quantum critical point in high-$T_c$ superconductors, which is ``hidden'' by the superconducting phase. Notwithstanding, one may uncover new scaling relations that can be tested experimentally. It seems likely that a realistic description will be in terms of a fluid with broken Galilean invariance (rather than Lorentzian), as we proposed for the normal phase in \cite{Hoyos:2013eza}.

We hope to address these questions and others in the future.

%%%%%%%%%%%%%%%%%%%%%%%%%%%%%%%%%%%%%%%%%%%%%%%%%%%%%%%%%%%
\section*{Acknowledgements}
%%%%%%%%%%%%%%%%%%%%%%%%%%%%%%%%%%%%%%%%%%%%%%%%%%%%%

We would like to thank J.~Bhattacharya, S. Chapman, J.~de Boer, K.~Jensen, E.~Kiritsis, R.~Loganayagam, R.~Meyer, G.~ Policastro, M.~Rangamani, S.~Sugimoto, A.~Yarom and P.~Yi for discussions and comments. 
This work is supported in part by the Israeli Science Foundation Center
of Excellence, and by the I-CORE program of Planning and Budgeting Committee and the Israel Science Foundation (grant number 1937/12).
BSK is grateful for warm hospitality and various discussions with the members of the KIAS, Seoul, and Kavli IPMU, Kashiwa.  

\appendix

%%%%%%%%%%%%%%%%%%%%%%%%%%%%%%%%%%%%%%%%%%%%%%%%%%%%%%%%%%%%%%%%%%%%%%%%%
\section{Variations of metric and curvature with respect to vierbein}\label{sec:vierbeins}
%%%%%%%%%%%%%%%%%%%%%%%%%%%%%%%%%%%%%%%%%%%%%%%%%%%%%%%%%%%%%%%%%%%%%%%%%%

We will use the following formulas:
\begin{itemize}
\item Variation of the metric
\begin{equation}
\frac{\delta g^{\sigma\rho}}{\delta e_\alpha^{\ c}}=-g^{\sigma\lambda}g^{\tau\rho}\frac{\delta g_{\lambda \tau}}{\delta e_\alpha^{\ c}}.
\end{equation}
\begin{equation}
\frac{\delta g_{\rho\nu}}{\delta e_\alpha^{\ c}}=\eta_{ac}(\delta_\rho^\alpha e_\nu^{ \ a}+\delta_\nu^\alpha e_\rho^{\ a}).
\end{equation}
\item Variation of the Christoffel symbol
\begin{equation}
\delta \Gamma^\sigma_{\ \mu\nu}=-g^{\sigma\lambda}\Gamma^\tau_{\ \mu\nu}\frac{\delta g_{\lambda \tau}}{\delta e_\alpha^{\ c}}\delta e_\alpha^{\ c}+g^{\sigma\lambda}\delta \Gamma_{\lambda\mu\nu}.
\end{equation}
Using that
\begin{align}
\notag \nabla_\mu \frac{\delta g_{\rho\nu}}{\delta e_\alpha^{\ c}}+\nabla_\nu \frac{\delta g_{\rho\mu}}{\delta e_\alpha^{\ c}}-\nabla_\rho \frac{\delta g_{\mu\nu}}{\delta e_\alpha^{\ c}}& =\partial_\mu \frac{\delta g_{\rho\nu}}{\delta e_\alpha^{\ c}}+\partial_\nu \frac{\delta g_{\rho\mu}}{\delta e_\alpha^{\ c}}-\partial_\rho \frac{\delta g_{\mu\nu}}{\delta e_\alpha^{\ c}}\\-2 \Gamma^\tau_{\ \mu\nu}\frac{\delta g_{\rho\tau }}{\delta e_\alpha^{\ c}}+2\eta_{ac}\Gamma_{\mu\nu}^\alpha e_\rho^{\ a}.
\end{align}
and
\begin{equation}
\nabla_\mu \frac{\delta g_{\rho\nu}}{\delta e_\alpha^{\ c}}=-\eta_{ac}\frac{\delta g_{\rho\nu}}{\delta e_\alpha^{\ b}}\omega_\mu^{ab},
\end{equation}
where the spin connection is
\begin{equation}
\omega_\mu^{ab}=\eta^{ac}e_c^\beta\nabla_\mu e_\beta^b.
\end{equation}
We have used that
\begin{equation}
\nabla_\mu e_\alpha^{\ c}+\eta_{ab}\omega_\mu^{ca}e_\alpha^{\ b}=0.
\end{equation}

One can show 
\begin{equation}
\delta \Gamma^\sigma_{\ \mu\nu}=\frac{1}{2}g^{\sigma\rho}\left[ \delta_\mu^\beta \frac{\delta g_{\rho\nu}}{\delta e_\alpha^{\ c}}+\delta_\nu^\beta \frac{\delta g_{\rho\mu}}{\delta e_\alpha^{\ c}}-\delta_\rho^\beta \frac{\delta g_{\mu\nu}}{\delta e_\alpha^{\ c}}\right]\cD_\beta\delta e_\alpha^{\ c}.
\end{equation}
Where
\begin{equation}
\cD_\beta  \delta e_\alpha^{\ c} =\nabla_\beta \delta e_\alpha^{\ c}-\eta_{ab}\omega_\beta^{ac}\delta e_\alpha^{\ b}.
\end{equation}
Then,
\begin{equation}
\delta \Gamma^\sigma_{\ \mu\nu}=F^{\sigma\alpha\beta}_{\ \ \ \mu\nu\,c}\cD_\beta\delta e_\alpha^{\ c},
\end{equation}
where
\begin{equation}
F^{\sigma\alpha\beta}_{\ \ \ \mu\nu\,c}=\eta_{ac}\left[g^{\sigma\alpha}e_{(\nu}^{\ a} \delta_{\mu)}^\beta-g^{\sigma\beta}e_{(\nu}^{\ a} \delta_{\mu)}^\alpha \right]+e_c^{\ \sigma}\delta_{(\nu}^\alpha \delta_{\mu)}^\beta\,.
\end{equation}
\item Variation of Ricci tensor
\begin{equation}
\delta R_{\mu\nu}=\nabla_\mu \delta \Gamma^\rho_{\ \rho\nu}-\nabla_\rho \delta\Gamma^\rho_{\ \mu\nu}.
\end{equation}
From the variation of the Christoffel symbol one gets
\begin{equation}
\delta R_{\mu\nu}=F^{\rho\alpha\beta}_{\ \ \  \rho\nu\,c}\nabla_\mu\cD_\beta\delta e_\alpha^{\ c}-F^{\rho\alpha\beta}_{\ \ \  \mu\nu\,c}\nabla_\rho\cD_\beta\delta e_\alpha^{\ c}+\cdots\,.
\end{equation}
The dots are terms proportional to derivatives of the background vierbein (they should give contributions that make the expression covariant $\nabla_\mu\to \cD_\mu$), they will vanish in flat space.

If we contract with $t^\mu$:
\begin{equation}
\delta R_{\mu\nu} t^\mu t^\nu=(e_c^{\ \alpha}\nabla_\parallel t^\beta\cD_\beta-t_c\nabla^\alpha t^\beta\cD_\beta+t_c t^\alpha\nabla^\beta\cD_\beta-t^\alpha \nabla_c t^\beta\cD_\beta)\delta e_\alpha^{ c}+\cdots.
\end{equation}
Contracting with $P_t^{\mu\nu}$:
\begin{equation}
\delta R_{\mu\nu} P_t^{\mu\nu}=(e_c^{\ \alpha}P_t^{\beta\gamma}\nabla_\beta\cD_\gamma-P_{t\,c}^{\ \beta}\nabla^\alpha \cD_\beta+ P_{t\,c}^{\ \alpha}\nabla^\beta\cD_\beta-P_t^{\alpha\beta} \nabla_c\cD_\beta)\delta e_\alpha^{ c}+\cdots.
\end{equation}

It is convenient to form the combinations
\begin{align}  \label{Rcombinations}
R_\parallel &=-\frac{1}{d-1}\left(R_{\mu\nu} t^\mu t^\nu+\frac{1}{d}R_{\mu\nu}P_t^{\mu\nu}\right),\\
R_\perp &=-\frac{1}{2(d-1)}\left(R_{\mu\nu} t^\mu t^\nu +R_{\mu\nu}P_t^{\mu\nu}\right).
\end{align}

\end{itemize}

%%%%%%%%%%%%%%%%%%%%%%%%%%%%%%%%%%%%%%%%%%%%%%%%%%%%%%%%%%%%%%%%%%%%%%
\section{Thermal correlators in free theories}\label{sec:thermal}
%%%%%%%%%%%%%%%%%%%%%%%%%%%%%%%%%%%%%%%%%%%%%%%%%%%%%%%%%%%%%%%%%%%%%%

The Euclidean correlator of the scalar field is
\begin{equation}
\vev{\phi \phi}(i\omega_n,\vec{q})=\frac{1}{(i\omega_n)^2-Q^2}.
\end{equation}
Where $Q^2=q^2$ in the relativistic theory and $Q^2=\frac{\kappa}{2}(q^2)^2$ for the $z=2$ free scalar theory. In the following we will generalize this expression to arbitrary $z$ and number of dimensions. Even though for arbitrary $z$ there is no local action, the generalization of these results to arbitrary $z$ can be done by replacing  the terms $\sim \kappa (q^2)^2\to \kappa (q^2)^z$ and the factor $q^2$ in the $T^{i0}$ component by $(q^2)^{z-1}$.

\subsection{Relativistic theory:}

\begin{itemize}
\item The one-point function is
\begin{equation}
\vev{T_{00}}=-\frac{1}{2\beta}\sum_n \int \frac{d^d q}{(2\pi)^d} \frac{(i\omega_n)^2+q^2}{(i\omega_n)^2-q^2}.
\end{equation}
The sums over Matusbara frequencies gives
\begin{equation}
\frac{1}{\beta}\sum_n \frac{(i\omega_n)^2}{(i\omega_n)^2-q^2}=-\frac{q}{2}(1+2 n_B(q)),
\end{equation}
and
\begin{equation}
\frac{1}{\beta}\sum_n \frac{1}{(i\omega_n)^2-q^2}=-\frac{1}{2q}(1+2 n_B(q)),
\end{equation}
where 
\begin{equation}
n_B(q)=f_B(\beta q)=\frac{1}{e^{\beta q}-1}.
\end{equation}
When we add the two contributions there is a temperature-independent part which is divergent and we should substract and a temperature dependent part, which is
\begin{equation}
\vev{T_{00}}= \int \frac{d^d q}{(2\pi)^d} q n_B(q)=\frac{V(S^{d-1})}{(2\pi)^d}\int_0^\infty dq\, q^d n_B(q)= \frac{V(S^{d-1})}{(2\pi)^d} T^{d+1} \int_0^\infty dx\, x^d f_B(x).
\end{equation}
For the stress tensor we get that the temperature-dependent part is
\begin{equation}
\vev{T_{ij}}=\frac{\delta_{ij}}{d}\vev{T_{00}}.
\end{equation}

\item The two-point function is
\begin{equation}
\vev{T_{0i}T_{0j}}=\frac{1}{\beta}\sum_n \int \frac{d^d q}{(2\pi)^d} \frac{(i\omega_n)^2 q_i q_j}{((i\omega_n)^2-q^2)^2}.
\end{equation}
We will use that inside the integral we can substitute
\begin{equation}
q_i q_j \longrightarrow \frac{1}{d}\delta_{ij} q^2.
\end{equation}
The sum over Matusbara frequencies we will need is
\begin{equation}
\frac{1}{\beta}\sum_n \frac{(i\omega_n)^2}{((i\omega_n)^2-q^2)^2}=-\frac{1}{2}\left( \frac{1}{2q}(1+2 n_B(q))+n_B'(q)\right).
\end{equation}
As before, there is a temperature independent term that we should subtract and the remaining contribution is
\begin{equation}
\vev{T_{0i}T_{0j}}=-\frac{V(S^{d-1})}{(2\pi)^d}\frac{\delta_{ij} T^{d+1}}{2d}\int_0^\infty dx\, x^{d+1}\left(\frac{1}{x}f_B(x)+f_B'(x)\right).
\end{equation}
We can use integration by parts on the second term
\begin{equation}
\int_0^\infty dx\, x^{d+1} f_B'(x)=-(d+1)\int_0^\infty dx\, x^d f_B(x).
\end{equation}
Then, the two-point function becomes
\begin{equation}
\vev{T_{0i}T_{0j}}=\frac{1}{2}\delta_{ij} \frac{V(S^{d-1})}{(2\pi)^d} T^{d+1}\int_0^\infty dx\, x^d f_B(x)=\frac{1}{2}\vev{T^{00}}\delta_{ij}.
\end{equation}
The relation found from hydrodynamics in Minwalla is indeed satisfied.

\end{itemize}

\subsection{Lifshitz theory:} 

We will compute the one- and two- point functions

\begin{itemize}
\item The one-point function is
\begin{equation}
\vev{T_{00}}=-\frac{1}{2\beta}\sum_n \int \frac{d^d q}{(2\pi)^d} \frac{(i\omega_n)^2+Q^2}{(i\omega_n)^2-Q^2}.
\end{equation}
Where $Q^2\equiv \frac{\kappa}{z} (q^2)^z$.
The sums over Matusbara frequencies gives
\begin{equation}
\frac{1}{\beta}\sum_n \frac{(i\omega_n)^2}{(i\omega_n)^2-Q^2}=-\frac{Q}{2}(1+2 n_B(Q)),
\end{equation}
and
\begin{equation}
\frac{1}{\beta}\sum_n \frac{1}{(i\omega_n)^2-Q^2}=-\frac{1}{2Q}(1+2 n_B(Q)).
\end{equation}
When we add the two contributions there is a temperature-independent part which is divergent and we should substract and a temperature dependent part, which is
\begin{align}
\notag \vev{T_{00}} &= \int \frac{d^d q}{(2\pi)^d} Q n_B(Q)=\frac{V(S^{d-1})}{(2\pi)^d}\int_0^\infty dq\, q^{d-1} Q n_B(Q)\\
&= \left(\frac{z}{\kappa}\right)^{\frac{d}{2z}}\frac{V(S^{d-1})}{(2\pi)^d} T^{\frac{d+z}{z}} \int_0^\infty dx\, x^{d-1+z} f_B(x^z).
\end{align}
For $z=2$ and $\kappa/z=1$ we get
\begin{equation}
\vev{T^{00}}=\frac{V(S^{d-1})}{(2\pi)^d} T^{\frac{d+2}{2}} \int_0^\infty dx\, x^{d+1} f_B(x^2).
\end{equation}
For the stress tensor we get that the temperature-dependent part is
\begin{equation}
\vev{T_{ij}}=\frac{z}{d}\delta_{ij}\vev{T_{00}}.
\end{equation}

\item The two-point function is
\begin{equation}
\vev{T_{<0i>}T_{<0j>}}_c=\frac{1}{\beta}\left(\frac{\kappa}{z}\right)^{c/z}\sum_n \int \frac{d^d q}{(2\pi)^d} \frac{(i\omega_n)^2 q_i q_j Q^{\frac{2c(z-1)}{z}}}{((i\omega_n)^2-Q^2)^2}.
\end{equation}
Where we use the notation for $c=0,1,2$
\begin{equation}
\vev{T_{<0i>}T_{<0j>}}_0=\vev{T_{0i}T_{0j}},\  \vev{T_{<0i>}T_{<0j>}}_1=\vev{T_{0i}T_{j0}},\  \vev{T_{<0i>}T_{<0j>}}_2=\vev{T_{i0}T_{j0}}.
\end{equation}
The sum over Matusbara frequencies we will need is
\begin{equation}
\frac{1}{\beta}\sum_n \frac{(i\omega_n)^2}{((i\omega_n)^2-Q^2)^2}=-\frac{1}{2}\left( \frac{1}{2Q}(1+2 n_B(Q))+n_B'(Q)\right).
\end{equation}
As before, there is a temperature independent term that we should subtract and the remaining contribution is
\begin{align}
\notag \vev{T_{<0i>}T_{<0j>}}_c &=-\left(\frac{z}{\kappa}\right)^{\frac{d+2-2c}{2z}} \frac{V(S^{d-1})}{(2\pi)^d}\frac{\delta_{ij}}{2d} T^{\frac{d+2-z+2c(z-1)}{z}}\\
&\times \int_0^\infty dx\, x^{d+1+2c(z-1)}\left(\frac{1}{x^z}f_B(x^z)+f_B'(x^z)\right).
\end{align}
We will use that
\begin{equation}
f_B'(x^z)=\frac{1}{\partial (x^z)/\partial x}\frac{\partial}{\partial x}f_B(x^z)=\frac{1}{z x^{z-1}}\frac{\partial}{\partial x}f_B(x^z).
\end{equation}
We can use integration by parts on the second term 
\begin{equation}
\int_0^\infty dx\, \frac{1}{z} x^{d+2-z+2c (z-1)} \partial_x f_B(x^z)=-\frac{(d+2-z+2c( z-1))}{z}\int_0^\infty dx\, x^{d+1-z+2c( z-1)} f_B(x^z).
\end{equation}
Then, the two-point function becomes
\begin{align}
\notag \vev{T_{<0i>}T_{<0j>}}_c &=\frac{(d+2-2 z+ 2 c(z-1))}{2 z d}\delta_{ij}\left(\frac{z}{\kappa}\right)^{\frac{d+2-2 c}{2z}} \frac{V(S^{d-1})}{(2\pi)^d} T^{\frac{d+2-z+2 c( z-1)}{z}}\\
&\times \int_0^\infty dx\, x^{d+1-z+2 c (z-1)} f_B(x^z).
\end{align}
For $c=1$ 
\begin{equation}
\vev{T_{0i}T_{j0}}=\frac{1}{2 z}\delta_{ij}\left(\frac{z}{\kappa}\right)^{\frac{d}{2z}} \frac{V(S^{d-1})}{(2\pi)^d} T^{\frac{d+z}{z}}\int_0^\infty dx\, x^{d+z-1} f_B(x^z).
\end{equation}
Therefore we have the relation
\begin{equation}
\vev{T_{0i}T_{j0}}=\vev{T_{i0}T_{0j}}=\frac{1}{2z} \delta_{ij} \vev{T_{00}},
\end{equation}
which is analogous to the relativistic formula but with a different factor.

\end{itemize}

%%%%%%%%%%%%%%%%%%%%%%%%%%%%%%%%%%%%%%%%%%%%%%%%%%%%%%%%%%%
\section{Curvature tensors in Lifshitz}\label{sec:curvature}
%%%%%%%%%%%%%%%%%%%%%%%%%%%%%%%%%%%%%%%%%%%%%%%%%%%%%%%%%%%

In the original coordinates, the non-zero Christoffel symbols are
\begin{equation}
\Gamma^r_{rr}=\frac{1}{2}g^{rr}\partial_r g_{rr}=\frac{1}{2}\frac{F}{H^2}\left(\frac{H^2}{F}\right)'=\frac{H'}{H}-\frac{1}{2}\frac{F'}{F},
\end{equation}
\begin{equation}
\Gamma^r_{\mu\nu}=-\frac{1}{2} g^{rr}\partial_r g_{\mu\nu}=\frac{1}{2}\frac{F F'}{H^2} \delta_\mu^0\delta_\nu^0-\frac{1}{2}\frac{F G'}{H^2} \delta^i_\mu\delta^j_\nu\delta_{ij},
\end{equation}
\begin{equation}
\Gamma^\mu_{\nu r}=\frac{1}{2}g^{\mu\alpha}\partial_rg_{\nu\alpha}=\frac{1}{2}\frac{F'}{F}\delta^\mu_0\delta^0_\nu+\frac{1}{2}\frac{G'}{G}\delta^\mu_i\delta_\nu^j \delta_j^i.
\end{equation}
Useful formulas:
\begin{equation}
\Gamma^\alpha_{r \alpha}=\frac{1}{2}\left[\left(\frac{F'}{F}\right)+d\left(\frac{G'}{G}\right) \right],
\end{equation}
\begin{equation}
\Gamma^\alpha_{r \beta}\Gamma^\beta_{r \alpha}=\frac{1}{4}\left[\left(\frac{F'}{F}\right)^2+d\left(\frac{G'}{G}\right)^2 \right].
\end{equation}

The non-zero components of the Ricci tensor are then
\begin{equation}
R_{rr}=-\partial_r\Gamma^\alpha_{\alpha r}+\Gamma^\alpha_{r \alpha}\Gamma^r_{r r}- \Gamma^\alpha_{r \beta} \Gamma^\beta_{r \alpha},
\end{equation}
\begin{equation}
R_{\mu\nu}=\partial_r \Gamma^r_{\mu\nu}+\Gamma^\alpha_{\alpha r}\Gamma^r_{\mu\nu}-\Gamma^r_{\mu\beta} \Gamma^\beta_{\nu r}-\Gamma^\alpha_{\mu r}\Gamma^r_{\nu\alpha}.
\end{equation}
There is a constribution in $R_{rr}$ that is singular at the horizon for the Lifshitz and black brane solutions
\begin{equation}
R_{rr}=-\frac{F''}{2 F}+\frac{F'}{2F} \left(\frac{H'}{H}-\frac{d
   G'}{2 G}\right)-\frac{d}{2}\left[\frac{G''}{G}+\frac{G'
   H'}{G H}+\frac{1}{2}\left(\frac{G'}{G}\right)^2\right].
\end{equation}

The components that enter in the horizon equation are
\begin{equation}
R_{tt}=\frac{F F''}{2 H^2}+\frac{d F F' G'}{4 G
   H^2}-\frac{F F' H'}{2 H^3},
\end{equation}
\begin{equation}
R_{ij}=\delta_{ij}\left[-\frac{F' G'}{2 H^2}-\frac{F G''}{2 H^2}+\frac{F
   G' H'}{2 H^3}+\frac{(d-2)F G'^2}{4 G H^2}\right].
\end{equation}

The Ricci scalar is
\begin{equation}
R=-\frac{F''}{H^2}+\frac{F'}{H^2} \left(\frac{H'}{H}-\frac{d G'}{G}\right)+\frac{F}{H^2}
   \left(\frac{d G' H'}{G H}-\frac{d G''}{G}\right) 
   -\frac{F}{H^2} \frac{d(d-3) {G'}^2}{4 G^2}.
\end{equation}

\subsection{First order terms in the Einstein tensor}

The projection of the Einstein tensor to the horizon is
\begin{equation}
R_{\mu A}\ell^A=\frac{1}{G^{(d-1)/2}} R_{\mu \nu}S^\nu =\frac{1}{G^{d/2}} \tilde\nabla_\nu Q_\mu^{\ \nu}-\partial_\mu\theta.
\end{equation}
The second term is second order in derivatives and we will ignore it.

The first term is
\begin{equation}
\tilde\nabla_\nu Q_\mu^{\ \nu}=\partial_\nu Q_\mu^{\ \nu}-\frac{1}{2} Q^{\nu\rho}\partial_\mu\gamma_{\nu\rho},
\end{equation}
where $\gamma_{\nu\rho}=G P_{\nu\rho}$ at the horizon and
\begin{equation}
Q_\mu^{\ \nu}=G^{d/2}\left[ \Theta_\mu^{\ \nu}-\kappa \delta^\mu_\nu\right],
\end{equation}
with
\begin{equation}
\kappa=\Theta_\mu^{ \ \nu}u^\mu,
\end{equation}
and
\begin{equation}
\Theta_\mu^\nu=\nabla_\mu \ell^\nu=\partial_\mu u^\nu+\Gamma^\nu_{\mu\alpha}u^\alpha.
\end{equation}
To leading order in derivatives
\begin{equation}
\Gamma^\nu_{\mu\alpha}=\frac{F'}{2H}u^\nu u_\mu u_\alpha-\frac{G'}{2H}u^\nu P_{\mu\alpha}.
\end{equation}
Then,
\begin{equation}
\Theta_\mu^\nu=-\frac{F'}{2H}u^\nu u_\mu,
\end{equation}
which fixes
\begin{equation}
\kappa=\frac{F'}{2H},
\end{equation}
and
\begin{equation}
Q_\mu^{\ \nu}=-G^{d/2}\kappa P_\mu^{\ \nu}.
\end{equation}
We can use
\begin{equation}
Q^{\mu \nu}=-G^{d/2}\kappa P^{\mu \nu}.
\end{equation}
Putting all together, 
\begin{equation}
R_{\mu \nu} \ell^\nu=\kappa\left[\frac{d}{2G}\left(\partial_\mu G-P_\mu^\nu \partial_\nu G \right)-\partial_\nu P_{\ \mu}^\nu -P_{\ \mu}^\nu\frac{\partial_\nu\kappa}{\kappa}\right].
\end{equation}

%%%%%%%%%%%%%%%%%%%%%%%%%%%%%%%%
% bibliography
%%%%%%%%%%%%%%%%%%%%%%%%%%%%%%%%

\bibliographystyle{JHEP}   % bibtex style file JHEP
\bibliography{LifshitzBib}

\end{document}